\documentclass[aps,pra,groupedaddress,amsmath,amssymb,tightenlines,notitlepage,nofootinbib,onecolumn,superscriptaddress,11pt,showkeys]{revtex4-2}
\pdfoutput=1

\usepackage{bm,bbm}% 
\usepackage{braket}
\usepackage{amsthm,amsmath,amssymb}
\usepackage[normalem]{ulem}

\usepackage[utf8]{inputenc}
\usepackage[colorlinks=true, allcolors=MidnightBlue!70!black!70!TealBlue]{hyperref}

\usepackage[page,toc,titletoc,title]{appendix}
\usepackage[title]{appendix}
\usepackage[dvipsnames,svgnames]{xcolor}
\usepackage{amsmath,amssymb}
\usepackage{changepage,afterpage}
\usepackage{etruscan}
\usepackage{tikz}
\usepackage{enumerate}
\usepackage{mathtools}
\usepackage{array,booktabs,multirow}
\usepackage{siunitx}

\usepackage{newpxtext,newpxmath}

\newtheorem{theorem}{Theorem}

\theoremstyle{definition}

\newtheoremstyle{red}{}{}{\normalfont}{}{\color{red!80!black}\bfseries}{.}{ }{}
\theoremstyle{red}

\let\nc\newcommand

\nc{\NegW}{W_\tau}

\nc{\RW}{\Omega_\FF}

\DeclareMathOperator{\Tr}{Tr}

\nc{\id}{\mathbbm{1}}

\renewcommand{\O}{\mathcal{O}}

\newcommand{\DD}{\mathbb{D}}

\newcommand{\FF}{\mathcal{F}}

\nc{\Dmax}{D_{\max}}

\let\O\OO

\nc{\SEP}{\mathrm{SEP}}
\nc{\STAB}{\mathrm{STAB}}
\nc{\PPT}{\mathrm{PPT}}
\nc{\PPTP}{\mathrm{PPTP}}
\nc{\SEPP}{\mathrm{SEPP}}
\nc{\SP}{\mathrm{KP}}
\nc{\FP}{\mathrm{FP}}
\nc{\CPTP}{{\mathrm{CPTP}}}

\nc{\Op}{\mathcal{O}}

\nc{\idc}{\mathrm{id}}
\nc{\ve}{\varepsilon}

\nc{\Omax}{\O_{\mathrm{max}}}

\usepackage{stackengine,moresize}
\stackMath

\nc{\sminfty}{{\infty,\bullet}}

\usepackage[most,breakable]{tcolorbox}
\renewenvironment{boxed}[1]%
  {\expandafter\ifstrequal\expandafter{#1}{orange}{\begin{tcolorbox}[colback=orange!5,colframe=orange!15,breakable,enhanced]}{\begin{tcolorbox}[colback=white,colframe=gray!10,breakable,enhanced]}}%
  {\end{tcolorbox}}

\nc{\regrob}{\DD_{s,\FF}^{\infty}}
\nc{\regrobs}{\DD_{s,\FF}^{\sminfty}}

\makeatletter 
\renewcommand\onecolumngrid{% 
\do@columngrid{one}{\@ne}%
\def\set@footnotewidth{\onecolumngrid}% <<<<<<<<<<<<<<<<
\def\footnoterule{\kern-6pt\hrule width 1.5in\kern6pt}%
}
\makeatother

\usepackage{fmtcount,refcount}

\let\oldproofname\proofname
\renewcommand{\proofname}{\rm\bf{\oldproofname}}

\usepackage{pifont}

\usepackage{newunicodechar}

\def\Tr{\operatorname{Tr}}

\def\id{\operatorname{id}}
\def\arcoth{\operatorname{arcoth}}
\def\1{\openone}

\def\d{\textbf{d}}
\def\keywords{\xdef\@thefnmark{}\@footnotetext}

\def\DD{D}

\allowdisplaybreaks
\begin{document}

\title{Pretty good measurement for bosonic Gaussian ensembles}
 
\author{Hemant K. Mishra}
\email{hemant.mishra@cornell.edu}
 \affiliation{School of Electrical and Computer Engineering, Cornell University, Ithaca, New York 14850, USA}
 \affiliation{Hearne Institute for Theoretical Physics, Department of Physics and Astronomy,
 and Center for Computation and Technology, Louisiana State University, Baton Rouge, Louisiana 70803, USA}

\author{Ludovico Lami}
\email{ludovico.lami@gmail.com}
\affiliation{Institute of Theoretical Physics and IQST, Universit\"{a}t Ulm, Albert-Einstein-Allee 11D-89069 Ulm, Germany}
\affiliation{QuSoft, Science Park 123, 1098 XG Amsterdam, the Netherlands}
\affiliation{Korteweg--de Vries Institute for Mathematics, University of Amsterdam, Science Park 105-107, 1098 XG Amsterdam, the Netherlands}
\affiliation{Institute for Theoretical Physics, University of Amsterdam, Science Park 904, 1098 XH Amsterdam, the Netherlands}

\author{Prabha Mandayam}
\email{prabhamd@physics.iitm.ac.in}
\affiliation{Center for Quantum Information, Communication and Computing, IIT Madras, Chennai 600036, India}
\affiliation{Department of Physics, Indian Institute of Technology Madras, Chennai 600036, India}

\author{Mark M. Wilde}
\email{wilde@cornell.edu}
 \affiliation{School of Electrical and Computer Engineering, Cornell University, Ithaca, New York 14850, USA}
 \affiliation{Hearne Institute for Theoretical Physics, Department of Physics and Astronomy,
 and Center for Computation and Technology, Louisiana State University, Baton Rouge, Louisiana 70803, USA}

%%%%%%%%%%%%%%%%%%%%%%%%%%%%%%%%%%%%%%%%%%%%%%%%%%%%%%%%%%%%%%%%%%
\begin{abstract}
The pretty good measurement is a fundamental analytical tool in quantum information theory,  giving a method for inferring the classical label that identifies a quantum state chosen probabilistically from an ensemble. 
% It is called “pretty good” because the error probability incurred in identifying the selected state is no more than twice that of the optimal measurement. 
Identifying and constructing the pretty good measurement for the class of bosonic Gaussian states is of immediate practical relevance in quantum information processing tasks. 
Holevo recently showed that the pretty good measurement for a bosonic Gaussian ensemble is a bosonic Gaussian measurement that attains the accessible information of the ensemble [{\it IEEE Trans.~Inf.~Theory, 66(9):5634–564, 2020}]. 
In this paper, we provide an alternate proof of Gaussianity of the pretty good measurement for a Gaussian ensemble of multimode bosonic states, with a focus on establishing an explicit and efficiently computable Gaussian description of the measurement. 
We also compute an explicit form of the mean square error of the pretty good measurement, which is relevant when using it for parameter estimation.

Generalizing the pretty good measurement is a quantum instrument, called the pretty good instrument. We prove that the post-measurement state of the pretty good instrument is a faithful Gaussian state if the input state is a faithful Gaussian state whose covariance matrix satisfies a certain condition. Combined with our previous finding for the pretty good measurement and provided that the same condition holds, it  follows that the expected output state is a faithful Gaussian state as well. In this case, we  compute an explicit Gaussian description of the post-measurement and expected output states.
Our findings imply that the pretty good instrument for bosonic Gaussian ensembles is no longer merely an analytical tool, but that it can also be implemented experimentally in quantum optics laboratories.
\end{abstract}

\maketitle


\bigskip

\noindent \textit{Dedicated to Alexander S.~Holevo on the occasion of his 80th birthday. Professor Holevo's numerous seminal contributions, from past \cite{holevo1973optimal} to present \cite{holevo2021accessible,holevo2021classical}, and his textbooks \cite{holevo1982,holevo2003statistical} have served as an inspiration and source of knowledge for generations of quantum information scientists.}

\newpage
\tableofcontents

%%%%%%%%%%%%%%%%%%%%%%%%%%%%%%%%%%%%%%%%%%%%%%%%%%%%%%%%%%%%%%%%%%
%%%%%%%%%%%%%%%%%%%%%%%%%%%%%%%%%%%%%%%%%%%%%%%%%%%%%%%%%%%%%%%%%%
%%%%%%%%%%%%%%%%%%%%%%%%%%%%%%%%%%%%%%%%%%%%%%%%%%%%%%%%%%%%%%%%%%

\section{Introduction}

Quantum measurement is a fundamental component of quantum mechanics, giving a method for guessing the classical label that identifies the state of a quantum system prepared from an ensemble of states with a known {\it a priori} probability distribution \cite{holevo2003statistical, wiseman2009quantum, busch2016quantum}. 
It has important applications in quantum communication \cite{helstrom1969quantum, barnett2009}, quantum key distribution \cite{bae2015quantum}, and quantum cryptography \cite{phoenix1995quantum, bouwmeester2000physics}; more generally, it is the basic way that we read out classical information encoded into quantum states.

The extraction of information from a finite-dimensional quantum system prepared in one of  finitely many quantum states has been well studied in the past several decades \cite{helstrom1969quantum, holevo1973optimal, holevo1982, hausladen1994pretty, walgate2000local, holevo2003statistical}.
A measurement for such systems, with an expected error probability not more than twice the optimal error probability \cite{barnum2002reversing},  was independently identified by several authors  \cite{Belavkin75, Belavkin75a,hausladen1994pretty, hughston1993complete} (see also \cite{H79}). Known as the {\it pretty good measurement} or the {\it square-root measurement}, it is a commonly considered measurement and analytical tool in quantum information theory \cite{hall1997quantum, moore2005distinguishing, bacon2005optimal, hayashi2008, mosca2008quantum, dall2011informational, holevo2012information, dalla2015optimality, iten2016pretty, cheng2022simple, studzinski2022square}. Recently, a quantum algorithm was proposed for implementing the pretty good measurement for an ensemble of quantum states in discrete-variable systems \cite{GLMQW22}. 
The pretty good measurement for any ensemble of quantum states has a canonical mathematical construction which also makes it a valid measurement for continuous-variable (CV) systems \cite{holevo2020gaussian, holevo2021classical, holevo2021accessible}. 
In his work  on the classical capacity and accessible information of a bosonic Gaussian ensemble \cite{holevo2020gaussian}, Holevo showed that the pretty good measurement of a Gaussian ensemble of multimode bosonic Gaussian states is a Gaussian measurement that attains the accessible information of the ensemble.

% For such systems, with any measurement of such systems is the expected error probability of correctly guessing the state of the system, we simply call it the {\it error probability} of the measurement. The measurement that achieves the minimum error probability is called the {\it optimal measurement}, or Helstrom measurement. The existence and necessary and sufficient conditions for an optimal measurement are known \cite{holevo1973optimal, yuen1975optimum}. However, except for a few special cases such as ensembles with two states \cite{levitin1994entropy} and two qubit states \cite{barnett2009quantum}, how to implement such a strategy in general is still an open problem.

In this paper, we provide a comprehensive study of the pretty good measurement for bosonic Gaussian ensembles, as well as its generalization, the pretty good instrument. We begin by furnishing an alternate proof of the Gaussianity of the pretty good measurement for a bosonic Gaussian ensemble, together with an explicit and efficiently computable Gaussian description of the measurement. We additionally compute an explicit form of the mean square error for this measurement. 
Next, we study the pretty good instrument \cite[Remark~14]{W15}, which is the quantum instrument generalizing the pretty good measurement of the ensemble. We prove that the post-measurement state of the pretty good instrument corresponding to a faithful Gaussian state, under a certain condition on the covariance matrix, is a faithful Gaussian state. Combined with our previous finding for the pretty good measurement and provided that the same condition holds, it  follows that the expected output state is a faithful Gaussian state as well. In this case, we also compute an explicit Gaussian description of both the post-measurement and expected output states.
With all of these findings in place, the pretty good instrument for multimode bosonic systems is no longer merely an analytical tool for theoretical derivations in quantum information, but it can also be implemented experimentally in quantum optics laboratories.

% In this paper, we consider a continuous ensemble of $n$-mode bosonic Gaussian states with Gaussian probability distribution over $\mathbb{R}^{2n}$, and study continuous analog of the pretty good measurement. Identifying and constructing the pretty good measurement for the class of bosonic Gaussian states is of immediate practical relevance in quantum information processing tasks. For instance, the accessible information of a bosonic Gaussian ensemble is attained by the pretty good measurement . We reproduce Holevo's \cite{holevo2020gaussian} finding that the pretty good measurement for any Gaussian ensemble, to which he refers to as observable {\it dual} to the ensemble, is a Gaussian measurement, and provide an explicit and efficiently computable form for it. 
% As such, this measurement is no longer merely an analytical tool for this case but can also be implemented experimentally in quantum optics laboratories.
%  We also compute the mean square error of the measurement.

Our paper is organized as follows.  Section~\ref{sec2} reviews some definitions and the basic theory of CV quantum systems and bosonic Gaussian states.
In Section~\ref{sec3},  we set up the notations of a bosonic Gaussian ensemble and state two of our main results, Theorem~\ref{thm1} on the Gaussianity of the pretty good measurement and its explicit form, and Theorem~\ref{thm2} on the mean square error of the measurement. We then study the pretty good instrument in Section~\ref{pgi}, stating our finding in Theorem~\ref{thm3}.
Appendices~\ref{app:golden} through \ref{invert2} contain detailed calculations and proofs that support the aforementioned results.

%%%%%%%%%%%%%%%%%%%%%%%%%%%%%%%%%%%%%%%%%%%%%%%%%%%%%%%%%%%%%%%%%%
%%%%%%%%%%%%%%%%%%%%%%%%%%%%%%%%%%%%%%%%%%%%%%%%%%%%%%%%%%%%%%%%%%
\section{Background}\label{sec2}
% In this section, we briefly recall some definitions and review basic theory of CV quantum systems and bosonic Gaussian states.

%%%%%%%%%%%%%%%%%%%%%%%%%%%%%%%%%%%%%%%%%%%%%%%%%%%%%%%%%%%%%%%%%%
\subsection{Quantum states and quantum channels}

A CV quantum system is associated with an infinite-dimensional, separable Hilbert space $\mathcal{H}$ over the complex field $\mathbb{C}$. A quantum state of the system is given by a  density operator $\rho$, which is a self-adjoint, positive semidefinite operator of unit trace acting on $\mathcal{H}$. Let $\mathcal{D}(\mathcal{H})$ denote the set of density operators and $\mathcal{B}(\mathcal{H})$ the space of bounded linear operators acting on $\mathcal{H}$. A quantum channel $\Phi$, between two CV systems represented by Hilbert spaces $\mathcal{H}$ and $\mathcal{K}$,  is a completely positive, trace-preserving linear map from $\mathcal{B}(\mathcal{H})$ to $\mathcal{B}(\mathcal{K})$. In particular, for all $\rho \in \mathcal{D}(\mathcal{H})$, we have $\Phi(\rho) \in \mathcal{D}(\mathcal{K})$.  

%%%%%%%%%%%%%%%%%%%%%%%%%%%%%%%%%%%%%%%%%%%%%%%%%%%%%%%%%%%%%%%%%%

\subsection{Quantum ensembles and measurements}

A quantum ensemble, denoted by $\{(p(x),\rho_x)\}_{x \in \Pi}$, consists of a probability density function $p(x)$ on the parameter space $\Pi$ with an underlying measure $\pi(\d x)$ and a measurable family of quantum states $\rho_x$ in $\mathcal{D}(\mathcal{H})$. For example, the parameter space $\Pi$ can be $\mathbb{R}^m$ associated with the Lebesgue measure.
A quantum measurement is given by a {\em positive operator-valued measure} (POVM). It is a family of self-adjoint, positive semidefinite operators $\{E_x\}_{x \in \Pi}$ satisfying
\begin{align}
    \int_{\Pi}\!\! \pi(\d x) E_x = \mathbb{I},
\end{align}
where $\mathbb{I}$ is the identity operator acting on the underlying Hilbert space. If the quantum system that is being measured is prepared in the state $\rho$, then the probability density $q(x)$ for the measurement outcome $x$ is given by the Born rule:
\begin{equation}
    q(x) = \Tr[E_x \rho], \quad x \in \Pi.
\end{equation}

The construction of the pretty good measurement associated with a given ensemble of quantum states is as follows. 
Let $\rho$ be the {\it average state} of the ensemble given by
\begin{align}
    \rho \coloneqq \int_{\Pi}\!\! \pi (\d x) p(x) \rho_x,
\end{align}
where the integral exists in the strong sense of the Banach space of trace-class operators \cite{holevo2020gaussian}.
The family of operators $\{E_x\}_{x \in \Pi}$ defined by
\begin{equation}%\label{new11eqn1}
    E_x \coloneqq p(x) \rho^{-\frac{1}{2}} \rho_x \rho^{-\frac{1}{2}}
\end{equation}
is the associated pretty good measurement and well defined if every $E_x$ is a bounded operator. In this case, the fact that $\{E_x\}_{x \in \Pi}$ is a POVM can be seen from the fact that each $E_x$ is positive semi-definite and from the following completeness condition:
\begin{equation}
    \int_{\Pi}\!\! \pi (\d x) \, E_x = \int_{\Pi}\!\! \pi (\d x)\, p(x) \rho^{-\frac{1}{2}} \rho_x \rho^{-\frac{1}{2}}
    = \rho^{-\frac{1}{2}} \left(\int_{\Pi}\!\! \pi(\d x)\,  p(x)  \rho_x\right) \rho^{-\frac{1}{2}}
    = \rho^{-\frac{1}{2}} \rho \rho^{-\frac{1}{2}} = \mathbb{I}.
\end{equation}

%%%%%%%%%%%%%%%%%%%%%%%%%%%%%%%%%%%%%%%%%%%%%%%%%%%%%%%%%%%%%%%%%%
\subsection{Quantum instrument}

Let $\{\mathcal{M}_x\}_{x\in \Pi}$ be a collection of completely positive and trace non-increasing maps, such that the map
$
    \int_{\Pi} \pi (\d x) \mathcal{M}_x 
$
is trace preserving, i.e., a quantum channel. The collection $\{\mathcal{M}_x\}_{x\in \Pi}$ is called a quantum instrument  and is the most general way of describing both the classical outcome of a measurement in addition to the post-measurement state \cite{davies1970operational,D76,holevo1982,Ozawa1984}. If the input state is $\tau$, then the probability density for observing the outcome $x$ is given by
$
    \Tr [\mathcal{M}_x(\tau)],
$
and the post-measurement state is given by
$
     \mathcal{M}_x(\tau)/\Tr [\mathcal{M}_x(\tau)]$.
% in which  allows for dis described as a map $\mathcal{M}$ defined by
% \begin{align}
%     \mathcal{M}(\cdot) \coloneqq \int_{\Pi}\!\! \pi (\d x) |x\rangle\!\langle x| \otimes \mathcal{M}_x(\cdot),
% \end{align}
% and for an input state $\tau$, the post-measurement state of the instrument is given by
% This means that an outcome of the quantum instrument is a classical-quantum state such that the classical register $\Pi$ stores the outcome of the measurement.
A quantum instrument is thus a generalization of a quantum measurement in the sense that it records both the measurement outcome and the post-measurement state. 
Associated with a POVM $\{E_x\}_{x\in\Pi}$ is a quantum instrument, given by the collection $\{K_x (\cdot) K_x^{\dag}\}_{x \in \Pi}$,
% \begin{align}
%     \mathcal{M}(\tau) &= \int_{\Pi}\!\! \pi (\d x) |x\rangle\!\langle x| \otimes K_x \tau K_x^\dagger,
% \end{align}
 where $E_x = K_x^{\dagger}K_x$ for all $x \in \Pi$.

%%%%%%%%%%%%%%%%%%%%%%%%%%%%%%%%%%%%%%%%%%%%%%%%%%%%%%%%%%%%%%%%%%
\subsection{Bosonic Gaussian states}

We briefly recall some mathematical definitions and basic results on quantum Gaussian states that will be useful in the development of the paper (see \cite{caruso2008multi, adesso2014continuous, S17} for reviews).
An $n$-mode CV quantum system is described by a density operator acting on a tensor-product Hilbert space
\begin{equation}
    \mathcal{H}=\bigotimes_{j=1}^n \mathcal{H}_j,
\end{equation}
with each $\mathcal{H}_j$ being an infinite-dimensional separable Hilbert space over $\mathbb{C}$. Associated with the $j$th mode are the position- and momentum-quadrature (self-adjoint) operators, denoted by $\hat{x}_j$ and $\hat{p}_j$, which satisfy the canonical commutation relations (CCR):
\begin{equation}
    [\hat{x}_j, \hat{p}_k]=i\hbar \delta_{j,k}, \quad \text{for all } j,k \in \{1,\ldots,n\}.
\end{equation}
 Here $[\hat{x}_j, \hat{p}_k]=\hat{x}_j\hat{p}_k- \hat{p}_k\hat{x}_j$ denotes the commutator of $\hat{x}_j$ and $\hat{p}_k$, $i$ the imaginary unit, $\hbar$ the reduced Planck's constant $h/2\pi$, and $\delta_{j,k}$ the Kronecker delta. In our treatment of CV systems, we set $\hbar=1$.  Let $\hat{r}$ denote the following vector of canonical operators:
\begin{equation}
    \hat{r} \coloneqq \left(\hat{r}_1,\ldots, \hat{r}_{2n} \right)^T \equiv \left(\hat{x}_1, \hat{p}_1,\ldots, \hat{x}_n, \hat{p}_n \right)^T.
\end{equation}
Let $\left[\hat{r}, \hat{r}^T \right]$ denote the $2n \times 2n$ matrix whose $(j,k)$th element is given by $[\hat{r}_j, \hat{r}_k]$.
The CCR can be represented in matrix form as 
\begin{equation}
    \left[\hat{r}, \hat{r}^T \right]= i \Omega, \quad \text{where} \quad \Omega= I_n \otimes \begin{bmatrix}0 & 1 \\ -1 & 0 \end{bmatrix} ,
\end{equation}
 and $I_n$ is the $n \times n$ identity matrix. 
An $n$-mode faithful Gaussian state $\rho$ can be written as
\begin{align}\label{neweqn10}
    \rho &=\dfrac{1}{Z_\rho} \exp\!\left[-\hat{H}_\rho\right], \\
    \hat{H}_\rho &\coloneqq \frac{1}{2} (\hat{r}-r_\rho)^T H_\rho (\hat{r}-r_\rho),\\
    Z_\rho &\coloneqq \sqrt{\det \left([V_\rho+i\Omega]/2 \right) },
\end{align}
where $\hat{H}_\rho$ is the quadratic Hamiltonian operator of the state, $H_\rho$ is a $2n \times 2n$ real positive-definite matrix that we refer to as the Hamiltonian matrix, $r_\rho \in \mathbb{R}^{2n}$ is equal to the mean vector of the state $r_\rho=\langle \hat{r} \rangle_\rho = \Tr\!\left[\hat{r}\rho \right]$, and $V_\rho$ is the $2n \times 2n$ positive-definite covariance matrix whose $(j,k)$th element is given by
\begin{equation}
    [V_\rho]_{j,k} = \Tr\!\left[\{(\hat{r}-r_\rho)_j,(\hat{r}-r_\rho)_k\} \rho \right].
\end{equation}
Here $\{\hat{a},\hat{b}\}=\hat{a} \hat{b}+\hat{b} \hat{a}$ denotes the anticommutator of two operators $\hat{a}$ and $\hat{b}$. The covariance matrix $V_\rho$ of a faithful Gaussian state $\rho$ satisfies the following uncertainty principle:
\begin{align}\label{faithful_condition}
    V_\rho +i \Omega > 0. 
\end{align}
%A Gaussian state is faithful (having full support) if $V_\rho +i \Omega > 0$.
We shall use the following relation from Lemma~$10$ of \cite{seshadreesan2018renyi} on the covariance matrix of a faithful Gaussian state:
\begin{equation}\label{new3eqn10}
    V_\rho - i \Omega = \sqrt{I+\left(V_\rho \Omega \right)^{-2}} V_\rho \left(V_\rho+i \Omega \right)^{-1} V_\rho \sqrt{I+\left(\Omega V_\rho \right)^{-2}}. 
\end{equation}
The matrices $H_\rho$ and $V_\rho$ are related as follows \cite{chen2005gaussian,holevo2011entropy}:
\begin{align}
    H_\rho &= 2i\Omega \arcoth\left(V_\rho i \Omega\right),\label{new1eqn1}\\
    V_\rho &=\coth \left(i\Omega H_\rho/2 \right)i \Omega \label{new1eqn2},
\end{align}
where
\begin{align}
    \coth(x) &= \dfrac{e^{x}+e^{-x}}{e^x-e^{-x}}, \\
    \arcoth(x) &= \dfrac{1}{2} \ln \left(\dfrac{x+1}{x-1} \right).
\end{align}
Let $W_\rho \coloneqq -V_\rho i \Omega$. The relations \eqref{new1eqn1} and \eqref{new1eqn2} give the following well-known Cayley transforms \cite{cayley1846quelques,cayley1895collected}:
\begin{align}
    \exp\!\left[ i\Omega H_\rho \right]&=\dfrac{W_\rho-I}{W_\rho+I},\label{new9eqn1}\\
    W_\rho &= \dfrac{I+\exp\!\left[i\Omega H_\rho\right]}{I-\exp\!\left[i\Omega H_\rho\right]},\label{new9eqn2}
\end{align}
where we have used the notation $\frac{A}{B}\coloneqq AB^{-1}$ for invertible matrices $A$ and $B$.
Define a unitary operator $\hat{D}(r)$ for $r \in \mathbb{R}^{2n}$  as
\begin{equation}
    \hat{D}(r) \coloneqq \exp\! \left[i r^T \Omega \hat{r} \right].
\end{equation}
This is also known as a Weyl displacement operator. Its inverse is given by $\hat{D}(r)^{\dagger}=\hat{D}(-r)$.
The displacement operator shifts the mean of a Gaussian state $\rho$ by $r$;  i.e., the mean vector of $\hat{D}(r)^\dagger \rho \hat{D}(r)$ is $r_\rho+r$.
 The covariance matrix of the state does not change by the action of a displacement operator.

%%%%%%%%%%%%%%%%%%%%%%%%%%%%%%%%%%%%%%%%%%%%%%%%%%%%%%%%%%%%%%%%%%
%%%%%%%%%%%%%%%%%%%%%%%%%%%%%%%%%%%%%%%%%%%%%%%%%%%%%%%%%%%%%%%%%%
\subsection{Gaussian measurements}

The measurement corresponding to the POVM 
\begin{equation}
\left\{\dfrac{1}{(2\pi)^n} \d y_m \ \hat{D}(-y_m) \rho_m \hat{D}(y_m)\right\}_{y_m \in \mathbb{R}^{2n}},
\end{equation}
where $\rho_m$ is a fixed $n$-mode Gaussian state with zero mean vector and a generic covariance matrix $V_m$, is a Gaussian measurement. As such, the following equality holds:
\begin{align}
    \mathbb{I} &= \dfrac{1}{(2\pi)^n}  \int_{\mathbb{R}^{2n}}\!\! \d y_m \ \hat{D}(-y_m) \rho_m \hat{D}(y_m).
\end{align}
 This measurement is also known as {general-dyne detection} \cite{S17}, as it represents a general form for a Gaussian measurement.

\section{Pretty good measurement for Gaussian states and mean square error}

\label{sec3}

% In this section, we review the continuous analog of the pretty good measurement for any ensemble of $n$-mode bosonic Gaussian states with Gaussian probability density over the parameter space $\mathbb{R}^{2n}$. We show that the pretty good measurement of the Gaussian ensemble is a Gaussian measurement, and explicitly compute its expression in terms of the mean vector and the covariance matrix of the average Gaussian state of the ensemble.

Let $\{(p(x), \rho_x)\}_{x \in \mathbb{R}^{2n}}$ be an ensemble of Gaussian states, such that the state 
 $\rho_x$ is defined as follows:
\begin{equation}
\label{neweqn4}
    \rho_x\coloneqq \hat{D}(-Lx) \rho_0 \hat{D}(Lx),
\end{equation}
where $\rho_0$ is a fixed $n$-mode faithful Gaussian state, and
$L$ is a $2n \times 2n$ real invertible matrix. Additionally, $p(x)$ is a Gaussian probability density function with a mean vector $\mu \in \mathbb{R}^{2n}$ and a $2n \times 2n$ real positive-definite covariance matrix $\Sigma$:
\begin{equation}\label{new2eqn6}
   p(x) = \dfrac{1}{(2\pi)^n\sqrt{\det \Sigma}} \exp\!\left[-\dfrac{1}{2} (x-\mu)^T \Sigma^{-1}(x-\mu)\right].
\end{equation}
We can say that the ensemble $\{(p(x), \rho_x)\}_{x \in \mathbb{R}^{2n}}$ is a quantum generalization of the normal location model, well known in classical estimation theory \cite[Example~1.1]{korostelev2011mathematical}.
Note that the mean vector~$r_{\rho_x}$ and covariance matrix $V_{\rho_x}$ of $\rho_x$ are given by
\begin{align}
    r_{\rho_x} = r_{\rho_0}+Lx, \quad  V_{\rho_x}=V_{\rho_0}.
\end{align}
The average state $\rho$ of the ensemble  
% \begin{equation}\label{neweqn6}
%     \rho \coloneqq \int_{\mathbb{R}^{2n}} \d x \ p(x) \rho_x ,
% \end{equation}
 is also a faithful Gaussian state, and its mean vector and covariance matrix are given by \cite[Section~5.3.2]{S17}:
\begin{align}\label{new14eqn1}
    r_{\rho} = r_{\rho_0}+L\mu,  \quad   V_{\rho} = V_{\rho_0}+2L \Sigma L^T.
\end{align}

% The pretty good measurement for the ensemble above is defined as the following POVM:
% \begin{equation}\label{new11eqn1}
%     E_x \coloneqq p(x) \rho^{-\frac{1}{2}} \rho_x \rho^{-\frac{1}{2}},
% \end{equation}
% clearly satisfying the completeness condition because
% \begin{equation}
%     \int_{\mathbb{R}^{2n}} \d x\, E_x = \int_{\mathbb{R}^{2n}} \d x\, p(x) \rho^{-\frac{1}{2}} \rho_x \rho^{-\frac{1}{2}}
%     = \rho^{-\frac{1}{2}} \left(\int_{\mathbb{R}^{2n}} \d x\,  p(x)  \rho_x\right) \rho^{-\frac{1}{2}}
%     = \rho^{-\frac{1}{2}} \rho \rho^{-\frac{1}{2}} = \mathbb{I}.
% \end{equation}

The following theorem states that the pretty good measurement associated with the Gaussian ensemble is a Gaussian measurement, and it also provides an explicit expression for it. 
\begin{boxed}{white}
\begin{theorem}\mbox{}\label{thm1}
% Let $\{p(x), \rho_x\}_{x \in \mathbb{R}^{2n}}$ be the Gaussian ensemble given by \eqref{neweqn4} and \eqref{new2eqn6}. 
Let $\{(p(x), \rho_x)\}_{x \in \mathbb{R}^{2n}}$ be a Gaussian ensemble such that $\rho_x = \hat{D}(-Lx) \rho_0 \hat{D}(Lx)$,
where $\rho_0$ is a fixed $n$-mode faithful Gaussian state,  
$L$ is a $2n \times 2n$ real invertible matrix, and $p(x)$ is a Gaussian probability density function with mean vector $\mu \in \mathbb{R}^{2n}$ and $2n \times 2n$ real positive-definite covariance matrix $\Sigma$.
The pretty good measurement $\{E_x\}_{x \in \mathbb{R}^{2n}}$ associated with the Gaussian ensemble is a Gaussian measurement. Its explicit Gaussian description is as follows:
\begin{align}\label{new3eqn15}
     E_x\d x = \dfrac{1}{(2\pi)^n}  \hat{D}(-y) \sigma \hat{D}(y) \d y,
\end{align}
where $\sigma$ is an $n$-mode faithful Gaussian state with zero mean vector and covariance matrix $V_{\sigma}$ given by 
\begin{align}\label{new10eqn3}
    V_{\sigma} = -V_\rho +\dfrac{1}{2} \sqrt{I+\left(V_\rho \Omega \right)^{-2}} V_\rho L^{-T}\Sigma^{-1}L^{-1} V_\rho \sqrt{I+\left(\Omega V_\rho \right)^{-2}},
\end{align}
and the measurement outcome $y$  is related to the ensemble parameter $x$ by
\begin{align}
    y &= r_\rho+ \dfrac{1}{2}\sqrt{I+\left(V_\rho \Omega\right)^{-2}} V_\rho L^{-T}\Sigma^{-1} (x-\mu). \label{new22eqn4}
\end{align}
\end{theorem}
\end{boxed}
\begin{proof}
    The proof is given in Appendix~\ref{app:main}.
\end{proof}

We now investigate the mean square error of the pretty good measurement for estimating the parameter of the ensemble. Let $X$ be a random variable over
$\mathbb{R}^{2n}$ with the probability density $p_X(x) = p(x)$, so that it represents the true value of the classical label of the ensemble. Also, let $\widetilde{X}$ be a random variable taking values in $\mathbb{R}^{2n}$ given by the outcomes of the pretty good measurement. The conditional probability density of $\widetilde{X}$ for given $X=x$ is given by the Born rule $p_{\widetilde{X}|X}(\tilde{x}|x)=\Tr[E_x \rho_{\tilde{x}}]=\Tr[p(x) \rho^{-\frac{1}{2}} \rho_x \rho^{-\frac{1}{2}} \rho_{\tilde{x}}]$. The mean square error of the pretty good measurement is defined as the expected value of $\|X-\widetilde{X}\|^2:$
\begin{equation}
    \begin{aligned}
    \mathbb{E}[\|X-\widetilde{X}\|^2]& \coloneqq \int_{\mathbb{R}^{2n}}\int_{\mathbb{R}^{2n}}\!\! \d x \d \tilde{x} \ \left \Vert x - \tilde{x} \right \Vert ^2 p_{X,\widetilde{X}} (x, \tilde{x})\\
    &=\int_{\mathbb{R}^{2n}}\int_{\mathbb{R}^{2n}}\!\! \d x \d \tilde{x} \ \left \Vert x - \tilde{x} \right \Vert ^2 p_X(x) p_{\widetilde{X}|X}(\tilde{x}|x) .\label{mse_formula}
\end{aligned}
\end{equation}
 In the following theorem,
we provide an exact expression for the mean square error of the pretty good measurement.
\begin{boxed}{white}
\begin{theorem}\mbox{}\label{thm2}
% Let $\{(p(x), \rho_x)\}_{x \in \mathbb{R}^{2n}}$ be a Gaussian ensemble given by $\rho_x = \hat{D}(-Lx) \rho_0 \hat{D}(Lx)$,
% where $\rho_0$ is a fixed $n$-mode faithful Gaussian state and  
% $L$ is a $2n \times 2n$ real invertible matrix, and $p(x)$ is a Gaussian probability density function with a mean vector $\mu \in \mathbb{R}^{2n}$ and a $2n \times 2n$ real positive-definite covariance matrix $\Sigma$.
The mean square error of the pretty good measurement associated with the Gaussian ensemble described in Theorem~\ref{thm1} is
\begin{align}\label{new10eqn6}
    \mathbb{E}[\|X-\widetilde{X}\|^2]=2 \Tr\!\left[\left(I-2 \Sigma L^T  V_\rho^{-1}\left(\sqrt{I+\left(V_\rho \Omega\right)^{-2}} \right)^{-1} L \right)\Sigma \right].
\end{align}
\end{theorem}
\end{boxed}

\begin{proof}
    See Appendix~\ref{app:mse} for a proof.
\end{proof}
% In general, an estimator of the ensemble parameter $X$ corresponding to the pretty good measurement is a function of $\widetilde{X}$.
% It is well known that the estimator $\mathbb{E}[X | \widetilde{X}]$, defined by
% \begin{align}
%     \mathbb{E}[X | \widetilde{X}=\tilde{x}] \coloneqq \int_{\mathbb{R}^{2n}}\!\! \d x \ x p_{X|\widetilde{X}}(x|\tilde{x}),
% \end{align}
% achieves the minimum mean square error (MMSE), i.e.,
% \begin{align}
%     \mathbb{E}[\|X-\mathbb{E}[X | \widetilde{X}]\|^2]= \inf_{\hat{X}} \mathbb{E}[\|X-\hat{X}\|^2],
% \end{align}
% where the infimum is taken over all the estimators $\hat{X}\equiv \hat{X}(\widetilde{X})$ of $X$. The estimator $\mathbb{E}[X | \widetilde{X}]$ is called the MMSE estimator.
% See \cite{kay1993fundamentals, van2004detection} for more details on estimation theory. 

%%%%%%%%%%%%%%%%%%%%%%%%%%%%%%%%%%%%%%%%%%%%%%%%%%%%%%%%%%%%%%%%%%
%%%%%%%%%%%%%%%%%%%%%%%%%%%%%%%%%%%%%%%%%%%%%%%%%%%%%%%%%%%%%%%%%%
\section{Pretty good instrument for Gaussian states}\label{pgi}

The quantum instrument associated with the pretty good measurement is called {\it the pretty good instrument} (see \cite[Remark~14]{W15}). For the Gaussian ensemble $\{(p(x), \rho_x)\}_{x \in \mathbb{R}^{2n}}$, it is defined as the following collection of completely positive, trace non-increasing maps:
\begin{align}
    \left\{\tau \mapsto  p(x) \rho_x^{1/2} \rho^{-1/2} \tau \rho^{-1/2} \rho_x^{1/2} \right\}_{x \in \mathbb{R}^{2n}}.
    \label{eq:pg-instr}
\end{align}
We emphasize that for every input state $\tau$, the operator $\rho_x^{1/2} \rho^{-1/2} \tau \rho^{-1/2} \rho_x^{1/2}$ is trace class. This follows because $\rho_x^{1/2} \rho^{-1/2}$ is a bounded operator, which is a consequence of the fact that the max-relative entropy of $\rho_x$ and $\rho$ is finite, since $V_{\rho_0} < V_{\rho}$. See Theorem~24 of \cite{seshadreesan2018renyi}.
We are interested in the probability density function $t(x)$ for observing outcome $x$
\begin{equation}
    t(x) = \Tr \!\left[p(x) \rho_x^{1/2} \rho^{-1/2} \tau \rho^{-1/2} \rho_x^{1/2} \right],
\end{equation}
the post-measurement state
\begin{align}
    \dfrac{p(x) \rho_x^{1/2} \rho^{-1/2} \tau \rho^{-1/2} \rho_x^{1/2}}{t(x)},
\end{align}
and the expected output state of the associated quantum channel 
\begin{align}
    \tau \mapsto \int_{\mathbb{R}^{2n}}\!\! \d x \ p(x) \rho_x^{1/2} \rho^{-1/2} \tau \rho^{-1/2} \rho_x^{1/2}.
    \label{eq:pg-channel}
\end{align}
The following theorem states that both the post-measurement state and the expected output state of the instrument are Gaussian if the input state $\tau$ is faithful Gaussian satisfying $V_\tau < V_\rho$. Its proof is given in Appendix~\ref{proof_pgi}.

\begin{boxed}{white}
\begin{theorem}\mbox{}\label{thm3}
For the pretty good instrument corresponding to the Gaussian ensemble described in Theorem~\ref{thm1}, if the input state $\tau$ is a faithful Gaussian state satisfying $V_\tau < V_\rho$, then the post-measurement state is a faithful Gaussian state and is given by
\begin{align}
    \dfrac{p(x) \rho_x^{1/2} \rho^{-1/2} \tau \rho^{-1/2} \rho_x^{1/2}}{\Tr \!\left[p(x) \rho_x^{1/2} \rho^{-1/2} \tau \rho^{-1/2} \rho_x^{1/2} \right]}=\hat{D}(-z) \rho_7 \hat{D}(z).
\end{align}
In the above, $\rho_7$ is a faithful Gaussian state with mean vector 
\begin{align}
    r_{\rho_7} &\coloneqq r_\rho + J_6J_5 (r_\tau - r_\rho),
\end{align}
and covariance matrix
\begin{align}
    V_{\rho_7} \coloneqq V_{\rho_0} - \sqrt{I+\left(V_{\rho_0} \Omega\right)^{-2}} V_{\rho_0} \left(V_5+V_{\rho_0}\right)^{-1} V_{\rho_0} \sqrt{I+\left(\Omega V_{\rho_0}\right)^{-2}},
\end{align}
where
\begin{align}
    V_5 &\coloneqq -V_\rho + \sqrt{I+\left(V_\rho \Omega\right)^{-2}} V_\rho \left(V_\rho-V_\tau\right)^{-1} V_\rho \sqrt{I+\left(\Omega V_\rho\right)^{-2}}, \label{new25eqn6} \\
    J_5 &\coloneqq \sqrt{I+\left(V_\rho \Omega\right)^{-2}}V_\rho\left(V_\rho-V_\tau\right)^{-1}, \label{new25eqn3} \\
    J_6 &\coloneqq\sqrt{I+\left(V_{\rho_0}\Omega\right)^{-2}}  V_{\rho_0}\left(V_5+V_{\rho_0}\right)^{-1}, \label{new25eqn4}\\
    J_7 &\coloneqq 2\left(I-J_6\right) L\Sigma L^T V_\rho^{-1}\left(\sqrt{I+\left(V_\rho \Omega\right)^{-2}}\right)^{-1} \label{new25eqn5}.
\end{align}
% \begin{multline}
%     r_{\rho_7} = (I-J_7)r_\rho + \sqrt{I+\left(V_{\rho_0}\Omega\right)^{-2}}  V_{\rho_0} \left[V_\rho \sqrt{I+\left(\Omega V_\rho\right)^{-2}} \right. \\ \left.- \left(V_\rho-V_\tau\right) V_\rho^{-1} \left(\sqrt{I+(V_\rho\Omega)^{-2}}\right)^{-1} (V_\rho-V_{\rho_0})\right]^{-1}
%     (r_\tau-r_\rho),
% \end{multline}
% and covariance matrix:
% \begin{multline}
%     V_{\rho_7} = V_{\rho_0} - \sqrt{I+\left(V_{\rho_0} \Omega\right)^{-2}} V_{\rho_0} \left(\sqrt{I+\left(V_\rho \Omega\right)^{-2}} V_\rho \left(V_\rho-V_\tau\right)^{-1} V_\rho \sqrt{I+\left(\Omega V_\rho\right)^{-2}} - (V_\rho-V_{\rho_0})\right)^{-1} \\ V_{\rho_0} \sqrt{I+\left(\Omega V_{\rho_0}\right)^{-2}},
% \end{multline}
% where
% \begin{multline}
%     J_7 = \left[I-\sqrt{I+\left(V_{\rho_0}\Omega\right)^{-2}} V_{\rho_0}\left(\sqrt{I+\left(V_\rho \Omega\right)^{-2}} V_\rho \left(V_\rho-V_\tau\right)^{-1} V_\rho \sqrt{I+\left(\Omega V_\rho\right)^{-2}} - (V_\rho-V_{\rho_0})\right)^{-1}\right] \\(V_\rho-V_{\rho_0}) V_\rho^{-1}\left(\sqrt{I+\left(V_\rho \Omega\right)^{-2}}\right)^{-1}.
% \end{multline}
The variable $z$ is related to $x$ by
\begin{align}
    z = J_7\sqrt{I+\left(V_\rho \Omega\right)^{-2}} V_\rho L^{-T}\Sigma^{-1}(x-\mu)/2.
\end{align}

In this case, the expected output state $\tilde{\tau}$ of the corresponding quantum channel \eqref{eq:pg-channel} is also a faithful Gaussian state, with  mean vector and covariance matrix given by
\begin{align}
    r_{\tilde{\tau}}& \coloneqq r_{\rho_7}+ J_7(r_\tau-r_\rho), \\
    V_{\tilde{\tau}}&\coloneqq V_{\rho_7}+J_7 \left[V_\tau + V_{\sigma} \right]J_7^T,
\end{align}
where $V_{\sigma}$ is given by \eqref{new10eqn3} in Theorem~\ref{thm1}.
\end{theorem}
\end{boxed}

\section{Conclusion}
One of the main findings of our work is a mathematically explicit Gaussian description of the pretty good measurement for an ensemble of multimode bosonic Gaussian states parameterized over $\mathbb{R}^{2n}$.
Furthermore, we have given a closed form of the mean square error for such a measurement. These results should be useful in experiments related to Bayesian quantum estimation tasks with Gaussian states, in which the goal is to estimate the vector $x$ in \eqref{neweqn4} by means of a measurement. Indeed, since the pretty good measurement in this case is a Gaussian measurement, the experimental demands of implementing this measurement are far less than if it were not.
Another finding of our work is a mathematically explicit Gaussian description of the post-measurement state, as well as the expected output state, of the pretty good instrument corresponding to a faithful Gaussian state $\tau$ under the condition $V_\tau < V_\rho$, where $\rho$ is the average state of the ensemble.

Going forward from here, it is an important open question to remove the need for the technical condition $V_{\tau} < V_{\rho}$ in Theorem~\ref{thm3} in order to establish that the pretty good instrument in \eqref{eq:pg-instr} is a Gaussian instrument. We suspect that this condition is not needed. It would also be interesting to make a more explicit connection between the findings presented here and the earlier results of \cite{lami2018approximate} for the Gaussian Petz recovery map, given that, in the finite-dimensional case, the pretty good instrument is known to be a special case of the Petz recovery map, as discussed in \cite[Remark~14]{W15}.

%%%%%%%%%%%%%%%%%%%%%%%%%%%%%%%%%%%%%%%%%%%%%%%%%%%%%%%%%%%%%%%%%%
%%%%%%%%%%%%%%%%%%%%%%%%%%%%%%%%%%%%%%%%%%%%%%%%%%%%%%%%%%%%%%%%%%
\section*{Acknowledgements}

HKM and MMW acknowledge support from the National Science Foundation under Grant No.~2304816.
The authors thank Tiju Cherian John, Aby Philip, Soorya Rethinasamy, Vishal Singh, Komal Malik, and Tanvi Jain for insightful discussions.

%%%%%%%%%%%%%%%%%%%%%%%%%%%%%%%%%%%%%%%%%%%%%%%%%%%%%%%%%%%%%%%%%%
%%%%%%%%%%%%%%%%%%%%%%%%%%%%%%%%%%%%%%%%%%%%%%%%%%%%%%%%%%%%%%%%%%
\bibliographystyle{unsrt}
\bibliography{pgm}

%%%%%%%%%%%%%%%%%%%%%%%%%%%%%%%%%%%%%%%%%%%%%%%%%%%%%%%%%%%%%%%%%%
%%%%%%%%%%%%%%%%%%%%%%%%%%%%%%%%%%%%%%%%%%%%%%%%%%%%%%%%%%%%%%%%%%
%%%%%%%%%%%%%%%%%%%%%%%%%%%%%%%%%%%%%%%%%%%%%%%%%%%%%%%%%%%%%%%%%%

%%%%%%%%%%%%%%%%%%%%%%%%%%%%%%%%%%%%%%%%%%%%%%%%%%%%%%%%%%%%%%%%%%
%%%%%%%%%%%%%%%%%%%%%%%%%%%%%%%%%%%%%%%%%%%%%%%%%%%%%%%%%%%%%%%%%%
% \appendix
% \addcontentsline{toc}{section}{Appendix} % Add the appendix text to the document TOC

\begin{appendices}
\section{The golden rule for manipulating exponential quadratic forms}

\label{app:golden}

We briefly recall the golden rule for manipulating products of exponential quadratic forms, which we use repeatedly in the paper. For more details, see \cite[Appendix~A]{lami2018approximate} and references therein.
Let $\hat{H}_j$ be any inhomogeneous quadratic operator of the form
\begin{align}\label{neweqn7}
    \hat{H}_j \coloneqq \dfrac{i}{2} \hat{r}^T \Omega X_j \hat{r}+i s_j^T \Omega \hat{r}+\dfrac{i}{2} a_j,
\end{align}
such that $X_j$ is a $2n \times 2n$ complex matrix with $\Omega X_j$ symmetric, $s_j \in \mathbb{C}^{2n}$, and $a_j \in \mathbb{C}$. Define a matrix $M_j(X_j,s_j,a_j)$ corresponding to the operator $\hat{H}_j$ as
\begin{equation}\label{neweqn8}
    M_j \equiv M_j(X_j,s_j,a_j) \coloneqq \begin{bmatrix}0 & s_j^T \Omega^T & a_j \\ 0 & X_j & s_j \\ 0 & 0 & 0 \end{bmatrix}.
\end{equation}
Its exponential is given by
\begin{equation}\label{new2eqn4}
    \exp[M_j] = \begin{bmatrix}1 &  \left( \dfrac{I-\exp[-X_j]}{X_j}s_j \right)^T \Omega^T & a_j+s_j^T \Omega \dfrac{X_j-\sinh X_j}{X_j^2} s_j \\ 0 & \exp[X_j] & \dfrac{\exp[X_j]-I}{X_j}s_j \\ 0 & 0 & 1 \end{bmatrix},
\end{equation}
 where $I$ is the $2n \times 2n$ identity matrix and $\sinh(x)=(e^x-e^{-x})/2$.
Given two operators $\hat{H}_1$ and $\hat{H}_2$, the operator $\hat{H}_3$ satisfying
\begin{equation}\label{neweqn9}
    \exp\!\left[\hat{H}_1\right] \exp\!\left[\hat{H}_2\right] = \exp\!\left[\hat{H}_3\right]
\end{equation}
lies in the Lie algebra generated by $\hat{H}_1$ and $\hat{H}_2$.
 The golden rule refers to the one-to-one correspondence between the operator $\hat{H}_3$ and its corresponding matrix $M_3$ satisfying
\begin{equation}\label{neweqn11}
      \exp[M_1] \exp[M_2] =\exp[M_3].
\end{equation}
It is easier to solve \eqref{neweqn11} for $M_3$ using the exponential form \eqref{new2eqn4} and  basic algebraic manipulations.

%%%%%%%%%%%%%%%%%%%%%%%%%%%%%%%%%%%%%%%%%%%%%%%%%%%%%%%%%%%%%%%%%%

%%%%%%%%%%%%%%%%%%%%%%%%%%%%%%%%%%%%%%%%%%%%%%%%%%%%%%%%%%%%%%%%%%
\section{The mean square formula for multidimensional Gaussian distributions }\label{gaussianformula}

Let $g: \mathbb{R}^m \to \mathbb{R}$ be a Gaussian probability density function with a mean vector $\eta \in \mathbb{R}^{m}$ and an $m \times m$ real positive-definite covariance matrix $\Gamma$:
\begin{equation}
   g(x) = \dfrac{1}{(2\pi)^{\frac{m}{2}}\sqrt{\det \Gamma}} \exp\!\left[-\dfrac{1}{2} (x-\eta)^T \Gamma^{-1}(x-\eta)\right].
\end{equation}
The density function $g$ satisfies the relation
\begin{align}\label{new1eqn7}
    \int_{\mathbb{R}^m}\!\!\d x\ \|x-y\|^2 g(x) &= \|\eta-y\|^2+\Tr \Gamma, \quad \text{for all } y \in \mathbb{R}^m.
\end{align}
We call \eqref{new1eqn7} the {\it mean square formula} for multidimensional Gaussian distributions. It can be easily proved using the following properties: for all $1 \leq i,j \leq m$,
\begin{align}
    \int_{\mathbb{R}^m}\!\!\d x\ x_i g(x) &= \eta_i,\\
    \int_{\mathbb{R}^m}\!\!\d x\ (x_i-\eta_i)(x_j-\eta_j) g(x) &= \Gamma_{ij}.
\end{align}
See Chapter~2 of \cite{gurland1966multidimensional} for a detailed treatment of integrals involving multidimensional Gaussian densities.

%%%%%%%%%%%%%%%%%%%%%%%%%%%%%%%%%%%%%%%%%%%%%%%%%%%%%%%%%%%%%%%%%%
\section{Proof of Theorem~\ref{thm1}}

\label{app:main}

Using the representation of Gaussian states in \eqref{neweqn10}, we can rewrite the pretty good measurement as
\begin{equation}\label{new4eqn2}
    E_x = p(x)Z_\rho Z_{\rho_0}^{-1}
\exp\!\left[\hat{H}_\rho/2\right] \exp\!\left[-\hat{H}_{\rho_x}\right] \exp\!\left[\hat{H}_\rho/2\right].
\end{equation}
Here we used the fact that the covariance matrix of $\rho_x$ is the same as that of $\rho_0$ so that $Z_{\rho_x}=Z_{\rho_0}$.
The golden rule, described in Appendix~\ref{app:golden}, guarantees that there exists an operator $\hat{H}_4$ of the form 
\begin{align}\label{neweqn12}
    \hat{H}_4 = \dfrac{i}{2} \hat{r}^T \Omega X_4 \hat{r}+i s_4^T \Omega \hat{r}+\dfrac{i}{2} a_4
\end{align}
that satisfies
\begin{align}\label{neweqn13}
    \exp\!\left[\hat{H}_\rho/2\right] \exp\!\left[-\hat{H}_{\rho_x}\right] \exp\!\left[\hat{H}_\rho/2\right]=\exp[\hat{H}_4].
\end{align}
The matrix $X_4$, vector $s_4$, and scalar $a_4$ in \eqref{neweqn12} can be obtained as follows.
Let $M_4$ be the matrix corresponding to the operator $\hat{H}_4$ given by \eqref{neweqn8}:
\begin{equation}
    M_4=\begin{bmatrix}0 & s_4^T \Omega^T & a_4 \\ 0 & X_4 & s_4 \\ 0 & 0 & 0 \end{bmatrix}.
\end{equation}
Write the operators $\hat{H}_\rho$ and $\hat{H}_{\rho_x}$ in the standard form \eqref{neweqn7}:
\begin{align}
    \hat{H}_\rho &= \dfrac{i}{2} \hat{r}^T \left(-i H_{\rho} \right)\hat{r} + i r_\rho^T \left(i H_\rho \right)\hat{r} + \dfrac{i}{2} \left(-i r_\rho^T H_\rho r_\rho \right), \\
    \hat{H}_{\rho_x} &= \dfrac{i}{2} \hat{r}^T \left(-i H_{\rho_0} \right)\hat{r} + i r_{\rho_x}^T \left(i H_{\rho_0} \right)\hat{r} + \dfrac{i}{2} \left(-i r_{\rho_x}^T H_{\rho_0} r_{\rho_x} \right),
\end{align}
and let $M_\rho$ and $M_{\rho_x}$ be the corresponding matrices given by \eqref{neweqn8}:
\begin{align}
    M_\rho &= \begin{bmatrix}0 & - i r_\rho^T H_\rho & -i r_\rho^T H_\rho r_\rho \\ 0 & i\Omega H_\rho & i\Omega H_\rho r_\rho \\ 0 & 0 & 0 \end{bmatrix},\\
    M_{\rho_x} &= \begin{bmatrix}0 &  -i r_{\rho_x}^T H_{\rho_0} & -i r_{\rho_x}^T H_{\rho_0} r_{\rho_x} \\ 0 & i\Omega H_{\rho_0} & i\Omega H_{\rho_0} r_{\rho_x} \\ 0 & 0 & 0 \end{bmatrix}.
\end{align}
The golden rule implies that the matrices $M_\rho, M_{\rho_x}$, and $M_4$ satisfy
\begin{align}\label{new25eqn1}
    \exp\!\left[M_\rho/2\right] \exp\!\left[-M_{\rho_x}\right] \exp\!\left[M_\rho/2\right]=\exp[M_4].
\end{align}
We know by \eqref{new2eqn4} that
\begin{align}
    \exp[M_\rho/2] &= \begin{bmatrix}1 & r_\rho^T\left(I-\exp[-i\Omega H_\rho/2] \right)^T \Omega^T & -r_\rho^T \Omega^T \sinh \left(i \Omega H_\rho/2 \right)r_\rho \\ 0 &  \exp[i\Omega H_\rho/2]  & \left(\exp[i\Omega H_\rho/2] - I \right) r_\rho \\ 0 & 0 & 1 \end{bmatrix}, \label{new3eqn8}\\
    \exp[-M_{\rho_x}] &= \begin{bmatrix}1 & r_{\rho_x}^T \left(I-\exp[i \Omega H_{\rho_0}] \right)^T \Omega^T & r_{\rho_x}^T \Omega^T \sinh \left(i\Omega H_{\rho_0} \right)r_{\rho_x} \\
    0 & \exp[-i \Omega H_{\rho_0}] & \left(\exp[-i\Omega H_{\rho_0}]-I \right) r_{\rho_x} \\ 0 & 0 & 1 \end{bmatrix}  \label{new3eqn9}, \\
    \exp[M_4] &= \begin{bmatrix}1 &  \left( \dfrac{I-\exp[-X_4]}{X_4}s_4 \right)^T \Omega^T & a_4+s_4^T \Omega \dfrac{X_4-\sinh X_4}{X_4^2} s_4 \\ 0 & \exp[X_4] & \dfrac{\exp[X_4]-I}{X_4}s_4 \\ 0 & 0 & 1 \end{bmatrix}. \label{new3eqn3}
\end{align}
Multiply the matrices in the left-hand side of \eqref{new25eqn1} using \eqref{new3eqn8} and \eqref{new3eqn9} and compare it with $\exp[M_4]$;  with some algebraic manipulations, we get
% \begin{align}\label{new3eqn1}
% \exp[M_4]=\exp[M_\rho/2]\exp[-M_{\rho_x}] \exp[M_\rho/2] &= \begin{bmatrix}
% 1 & s^T & \alpha \\ 0 & \Lambda & w \\ 0 & 0 & 1
% \end{bmatrix},
% \end{align}
% where
\begin{align}
    \exp[X_4] &= \exp[i\Omega H_\rho/2] \exp[-i \Omega H_{\rho_0}] \exp[i \Omega H_\rho/2],\label{new3eqn5} \\
    % \left( \dfrac{I-\exp[-X_4]}{X_4}s_4 \right)^T \Omega^T &= r_\rho^T \Omega^T \left(I-\Lambda \right)+ (r_{\rho_x}-r_{\rho})^T \Omega^T \left(I- \exp[-i \Omega H_{\rho_0}]\right)\exp[i\Omega H_\rho/2] ,\label{new3eqn4} \\
    \dfrac{\exp[X_4]-I}{X_4}s_4 & = \left( \exp[X_4]-I \right)r_\rho - \exp[i\Omega H_\rho/2] \left(I- \exp[-i\Omega H_{\rho_0}] \right)(r_{\rho_x}-r_{\rho}).\label{new3eqn7}
     % \alpha &\coloneqq (r_{\rho_x}-r_{\rho})^T \Omega \left[ \left(I-\exp[i\Omega H_{\rho_0}] \right)\exp[-i\Omega H_\rho/2]-\exp[i\Omega H_\rho/2]- \exp[-i \Omega H_{\rho_0}] \left(I-\exp[i\Omega H_\rho/2] \right) \right]r_\rho  \nonumber \\ 
     % &\hspace{0.5cm} + r_{\rho_x}^T \Omega^T \sinh \left(i\Omega H_{\rho_0} \right) r_{\rho_x}   + r_\rho^T \Omega \left(\Lambda r_\rho - \exp[-i\Omega H_{\rho_0}] r_{\rho_x} \right). \label{new3eqn6}
\end{align}
From \eqref{new3eqn5}, by applying Proposition~$6$ of \cite{seshadreesan2018renyi} twice, we  get $X_4=-i \Omega H_4$,
where $H_4 =2i \Omega \arcoth \left(V_4 i \Omega \right)$ and $V_4$ is a $2n \times 2n$ real symmetric matrix given by
\begin{align}
    V_4 &= -V_\rho +\sqrt{I+\left(V_\rho \Omega \right)^{-2}} V_\rho \left(V_\rho-V_{\rho_0} \right)^{-1} V_\rho \sqrt{I+\left(\Omega V_\rho \right)^{-2}} \label{new20eqn1} \\
    &=-V_\rho +\dfrac{1}{2} \sqrt{I+\left(V_\rho \Omega \right)^{-2}} V_\rho L^{-T}\Sigma^{-1}L^{-1} V_\rho \sqrt{I+\left(\Omega V_\rho \right)^{-2}}. \label{new4eqn1}
\end{align}
We used the relation $V_\rho-V_{\rho_0} =2L\Sigma L^T$ in the last equality.
Using the covariance matrix relation \eqref{new3eqn10} for $V_\rho$ in \eqref{new20eqn1}, we get
\begin{align}
    V_4+i\Omega &= -\left( V_\rho-i\Omega\right)  +\sqrt{I+\left(V_\rho \Omega \right)^{-2}} V_\rho \left(V_\rho-V_{\rho_0} \right)^{-1} V_\rho \sqrt{I+\left(\Omega V_\rho \right)^{-2}}\\
    &= \sqrt{I+\left(V_\rho \Omega \right)^{-2}} V_\rho \left[ \left(V_\rho-V_{\rho_0} \right)^{-1}-\left(V_\rho+i\Omega  \right)^{-1} \right]V_\rho \sqrt{I+\left(\Omega V_\rho \right)^{-2}} \label{neweqn3}\\
    &> 0.
\end{align}
The last inequality follows, since $V_{\rho_0}+i\Omega > 0$ implies $\left( V_\rho-V_{\rho_0}\right)^{-1}>\left( V_\rho+i \Omega \right)^{-1}$.
Therefore, $V_4$ is a legitimate covariance matrix of a faithful Gaussian state \cite{simon1994quantum}; we denote by $\rho_4$ the Gaussian state with zero mean vector and covariance matrix $V_{\rho_4}=V_4$.

Using the relations $X_4=-i\Omega H_4$ and $r_{\rho_x}-r_\rho=L(x-\mu)$ in \eqref{new3eqn7}, we get
\begin{align}\label{new5eqn2}
     s_4 = -i\Omega H_4 \left[ r_\rho + \left( \exp[-i\Omega H_4]-I \right)^{-1} \exp[i\Omega H_\rho/2] \left( \exp[-i\Omega H_{\rho_0}]-I \right)L(x-\mu) \right].
\end{align}
 Using the Cayley transform in  \eqref{new9eqn1}, we get
\begin{align}\label{new25eqn2}
    s_4 &= -i\Omega H_4 \left[ r_\rho + \left(W_{\rho_4}-I\right)  \exp[i\Omega H_\rho/2] \left(W_{\rho_0}-I\right)^{-1} L(x-\mu) \right].
\end{align}
By Corollary~4 and the function relation in Eq.~$(130)$ of \cite{seshadreesan2018renyi}, we get
\begin{align}
    \exp[i\Omega H_\rho/2] &= \sqrt{I-W_\rho^{-2}} \left(W_{\rho}+I\right)^{-1} W_\rho.
\end{align}
Substituting the above relation in \eqref{new25eqn2} gives
\begin{align}
     s_4  &   =  -i\Omega H_4 \left[ r_\rho + \left(W_{\rho_4}-I\right)  \sqrt{I-W_\rho^{-2}} \left(W_{\rho}+I\right)^{-1} W_\rho \left(W_{\rho_0}-I\right)^{-1} L(x-\mu)\right] \\
        &   = -i\Omega H_4 \left[ r_\rho + \left(W_{\rho_4}-I\right)  \sqrt{\left(I-W_\rho^{-1}\right)\left(I+W_\rho^{-1}\right)} \left(I+W_{\rho}^{-1}\right)^{-1} \left(W_{\rho_0}-I\right)^{-1} L(x-\mu)\right]\\
        &   = -i\Omega H_4 \left[ r_\rho + \left(W_{\rho_4}-I\right)  \sqrt{\left(I-W_\rho^{-1}\right)\left(I+W_\rho^{-1}\right)^{-1}} \left(W_{\rho_0}-I\right)^{-1} L(x-\mu)\right]\\
        &   = -i\Omega H_4 \left[ r_\rho + \left(W_{\rho_4}-I\right)  \sqrt{\left(I-W_\rho^{-1}\right)^{-1}\left(I+W_\rho^{-1}\right)^{-1}} \left(I-W_\rho^{-1}\right) \left(W_{\rho_0}-I\right)^{-1} L(x-\mu)\right]\\
        &   =-i\Omega H_4 \left[ r_\rho + \left(W_{\rho_4}-I\right)  \sqrt{\left(I-W_\rho^{-2}\right)^{-1}} \left(I-W_\rho^{-1}\right) \left(W_{\rho_0}-I\right)^{-1} L(x-\mu) \right]\label{new4eqn10}.
\end{align}
From \eqref{neweqn3} we get 
\begin{align}
    W_{\rho_4}-I &= \sqrt{I-W_\rho^{-2}} W_\rho \left[\left(W_\rho - W_{\rho_0}\right)^{-1}-\left(W_\rho -I\right)^{-1} \right]W_\rho \sqrt{I-W_\rho^{-2}} \\
    &= \sqrt{I-W_\rho^{-2}} W_\rho \left(W_\rho - W_{\rho_0}\right)^{-1} \left[\left(W_\rho -I\right)-\left(W_\rho - W_{\rho_0}\right) \right]\left(W_\rho -I\right)^{-1} W_\rho \sqrt{I-W_\rho^{-2}} \\
    &=\sqrt{I-W_\rho^{-2}} W_\rho \left(W_\rho - W_{\rho_0}\right)^{-1} \left(W_{\rho_0} -I\right) \left(W_\rho -I\right)^{-1} W_\rho \sqrt{I-W_\rho^{-2}}.
\end{align}
By substituting the above expression of $W_{\rho_4}-I$ in \eqref{new4eqn10}, we get
\begin{align}
     s_4 &=-i\Omega H_4 \left[ r_\rho + \sqrt{I-W_\rho^{-2}} W_\rho \left(W_\rho - W_{\rho_0}\right)^{-1} L(x-\mu) \right]\eqqcolon -i\Omega H_4 y,
\end{align}
where 
\begin{align}
    y &= r_\rho+J L(x-\mu), \label{new20eqn9}\\
    J &\coloneqq  \sqrt{I-W_\rho^{-2}} W_\rho \left(W_\rho - W_{\rho_0}\right)^{-1}\label{new6eqn1} =\sqrt{I+\left(V_\rho \Omega\right)^{-2}} V_\rho \left(V_\rho - V_{\rho_0}\right)^{-1}\\
    & = \sqrt{I+\left(V_\rho \Omega\right)^{-2}} V_\rho \left(2L\Sigma L^T\right)^{-1} \label{matrix_j_simplest_form}.
\end{align}
By substituting the values $X_4=-i\Omega H_4$ and $s_4=-i\Omega H_4 y$ in \eqref{neweqn12} and simplifying, we get
\begin{align}\label{new3eqn2}
\hat{H}_4 &=-\frac{1}{2} \left(\hat{r}-y \right)^T H_4 \left(\hat{r}-y\right) + \dfrac{1}{2} \left( ia_4+y^T H_4 y \right).
\end{align}
By substituting \eqref{new3eqn2} into \eqref{neweqn13} and then combining with \eqref{new4eqn2}, we get
\begin{align}
 E_x &=   p(x) Z_\rho Z_{\rho_0}^{-1} \exp\!\left[\dfrac{1}{2} \left( ia_4+y^T H_4 y \right) \right]
 \exp\!\left[-\frac{1}{2} \left(\hat{r}-y \right)^T H_4 \left(\hat{r}-y\right)\right]\\
 &=  \xi p(x) Z_\rho Z_{\rho_4} Z_{\rho_0}^{-1}
 \hat{D}(-y) \rho_4 \hat{D}(y), \label{new8eqn1}
\end{align}
where $\xi\coloneqq\exp[(ia_4 + y^TH_4 y)/2]$.

In the remainder of the proof, we will establish that $\xi = (2\pi)^{-n} (\det \Sigma)^{-1/2} p(x)^{-1}$, which will in turn allow us to conclude the final form of the Gaussian pretty good measurement. 
By the same arguments as in Proposition~$12$ of \cite{seshadreesan2018renyi}, applied to \eqref{new3eqn5},
we get
\begin{align}
    Z_{\rho_4} &=\sqrt{\det\left(\left[V_{\rho_4}+i \Omega\right]/2\right)} \\
    &=  \sqrt{\dfrac{\det \left(\left[ V_\rho +i\Omega \right]/2 \right)\det \left(\left[ V_{\rho_0} +i\Omega \right]/2 \right)}{\det \left(\left[ V_\rho - V_{\rho_0} \right]/2 \right)}} \\
    &= \sqrt{\dfrac{\det \left(\left[ V_\rho +i\Omega \right]/2 \right)\det \left(\left[ V_{\rho_0} +i\Omega \right]/2 \right)}{\det L^2 \det \Sigma}}\\
    &=\dfrac{Z_\rho Z_{\rho_0}}{\left|\det L\right| \sqrt{\det \Sigma}}.\label{new9eqn3}
\end{align}
Substituting the value of $Z_{\rho_4}$ into \eqref{new8eqn1} gives
\begin{equation}\label{new8eqn4}
    E_x =\dfrac{\xi p(x)  Z_\rho^2}{\left|\det L\right| \sqrt{\det \Sigma}} \hat{D}(-y) \rho_4 \hat{D}(y).
\end{equation}
Also, from \eqref{new6eqn1} we have
\begin{align}
    \det J 
    &= \det \sqrt{I+\left(V_\rho \Omega\right)^{-2}} \det V_\rho \det \left(V_\rho - V_{\rho_0}\right)^{-1} \\
    &=\det \sqrt{I-\left(V_\rho i \Omega\right)^{-2}} \det V_\rho \det \left(V_\rho - V_{\rho_0}\right)^{-1} \\
    &= \sqrt{\det \left[I-\left(V_\rho i \Omega\right)^{-1}\right] \det \left[I+\left(V_\rho i \Omega\right)^{-1}\right]} \det V_\rho \det \left(V_\rho - V_{\rho_0}\right)^{-1} \\
    &= \sqrt{ \det \left[V_\rho- i \Omega \right] \det V_\rho^{-1}  \det \left[V_\rho+ i \Omega\right] \det V_\rho^{-1}} \det V_\rho \det \left(V_\rho - V_{\rho_0}\right)^{-1} \\
    &= \sqrt{ \det \left[V_\rho+ i \Omega\right]  \det \left[V_\rho+ i \Omega\right]} \det \left(2L\Sigma L^T\right)^{-1} \\
    &= \det\left([V_\rho+ i\Omega]/2\right) \left(\det L^2 \det \Sigma \right)^{-1}     \\
    &= \dfrac{Z_\rho^2 }{ \det L^2 \det \Sigma},
\end{align}
which implies
\begin{align}
    \dfrac{Z_\rho^2 }{ \left|\det L\right| \sqrt{\det \Sigma}}=\left|\det [JL]\right|  \sqrt{\det \Sigma}. \label{new8eqn5}
\end{align}
From \eqref{new8eqn4} and \eqref{new8eqn5}, we get
\begin{align}\label{neweqn1}
    E_x =\xi p(x) \left|\det [JL]\right|  \sqrt{\det \Sigma} \ \hat{D}(-y) \rho_4 \hat{D}(y).
\end{align}
 We have $\Tr[E_x \rho]=\Tr[\rho^{-1/2} p(x) \rho_x \rho^{-1/2} \rho]=p(x)$. 
Using \eqref{neweqn1}, we thus get
\begin{align}
    p(x) &= \xi p(x) \left|\det [JL]\right|  \sqrt{\det \Sigma} \Tr[\hat{D}(-y) \rho_4 \hat{D}(y)\rho],
\end{align}
which implies
\begin{align}\label{new6eqn2}
\xi&=\dfrac{1 }{ \left|\det [JL]\right|  \sqrt{\det \Sigma} \Tr[\hat{D}(-y) \rho_4 \hat{D}(y) \rho]}.
\end{align}
By the overlap formula for Gaussian states  \cite[Eq.~(4.51)]{S17}, we get
\begin{align}\label{new6eqn3}
    \Tr[\hat{D}(-y) \rho_4 \hat{D}(y) \rho]&=\dfrac{1}{\sqrt{\det([V_{\rho_4} + V_\rho ]/2)}} \exp\! \left[-\frac{1}{2}(y-r_\rho)^T ([V_{\rho_4}+V_\rho]/2 )^{-1} (y-r_\rho) \right].
\end{align}
From \eqref{new6eqn2}, \eqref{new6eqn3}, and using the relation $y=r_\rho+JL(x-\mu)$, we thus get
\begin{align}
    \xi
        &=\dfrac{\sqrt{\det([V_{\rho_4} + V_\rho ]/2)}}{ \left|\det [JL]\right|  \sqrt{\det \Sigma}}\exp\! \left[\frac{1}{2}(x-\mu)^T \left(L^{-1}J^{-1}[V_{\rho_4}+V_\rho]J^{-T}L^{-T}/2 \right)^{-1} (x-\mu) \right] \\
        &=\dfrac{\sqrt{(\det [JL])^{-2}\det([V_{\rho_4} + V_\rho ]/2)}}{ \sqrt{\det \Sigma}}\exp\! \left[\frac{1}{2}(x-\mu)^T \left(L^{-1}J^{-1}[V_{\rho_4}+V_\rho]J^{-T}L^{-T}/2 \right)^{-1} (x-\mu) \right] \\
        &=\dfrac{\sqrt{\det(L^{-1}J^{-1}[V_{\rho_4} + V_\rho ]J^{-T}L^{-T}/2)}}{ \sqrt{\det \Sigma}}\exp\! \left[\frac{1}{2}(x-\mu)^T \left(L^{-1}J^{-1}[V_{\rho_4}+V_\rho]J^{-T}L^{-T}/2 \right)^{-1} (x-\mu) \right].\label{new24eqn2}
\end{align}
From \eqref{new4eqn1} and \eqref{new6eqn1} and the fact that $V_{\rho_4} = V_4$, we have 
\begin{align}
    \frac{1}{2}L^{-1}J^{-1}[V_{\rho_4}+V_\rho]J^{-T}L^{-T} %&= \frac{1}{2}L^{-1}J^{-1}[V_4+V_\rho]J^{-T}L^{-T} \\
    &=\frac{1}{2} L^{-1}J^{-1} \sqrt{I+\left(V_\rho \Omega \right)^{-2}} V_\rho \left(V_\rho-V_{\rho_0} \right)^{-1} V_\rho \sqrt{I+\left(\Omega V_\rho \right)^{-2}} J^{-T}L^{-T} \\
    &=\frac{1}{2}L^{-1} \left(V_\rho - V_{\rho_0}\right) L^{-T}\\
    &=\frac{1}{2}L^{-1} \left(2L\Sigma L^T\right) L^{-T}\\
    &=\Sigma.\label{new24eqn1}
\end{align}
Substituting \eqref{new24eqn1} into \eqref{new24eqn2} gives
\begin{align}
    \xi
    &= \exp\! \left[\frac{1}{2}(x-\mu)^T \Sigma^{-1} (x-\mu) \right]= (2\pi)^{-n} (\det \Sigma)^{-1/2} p(x)^{-1}.
\end{align}
By substituting the above value of $\xi$ into \eqref{neweqn1}, we get
\begin{align}\label{new20eqn8}
     E_x =\dfrac{\left|\det [JL]\right|}{(2\pi)^n} \hat{D}(-y) \rho_4 \hat{D}(y).
\end{align}
Recall from \eqref{new20eqn9} that $y=r_\rho + JL(x-\mu)$, which implies $\d y = \left|\det [JL]\right| \d x$. We thus get
\begin{align}
     E_x \d x =\dfrac{1}{(2\pi)^n} \hat{D}(-y) \rho_4 \hat{D}(y) \d y.
\end{align}
Finally, we make the substitution $\rho_4 \to \sigma$ to arrive at the precise statement given in Theorem~\ref{thm1}.

%%%%%%%%%%%%%%%%%%%%%%%%%%%%%%%%%%%%%%%%%%%%%%%%%%%%%%%%%%%%%%%%%%
%%%%%%%%%%%%%%%%%%%%%%%%%%%%%%%%%%%%%%%%%%%%%%%%%%%%%%%%%%%%%%%%%%
\section{Proof of Theorem~\ref{thm2}}
\label{app:mse}

Recall from \eqref{mse_formula} that, we have

\begin{align}
    \mathbb{E}[\|X-\widetilde{X}\|^2]&= \int_{\mathbb{R}^{2n}}\int_{\mathbb{R}^{2n}}\!\! \d x \d \tilde{x} \ \left \Vert x - \tilde{x} \right \Vert ^2 p(x) p_{\widetilde{X}|X}(\tilde{x}|x) \label{new4eqn6},
\end{align}
where $p_{\widetilde{X}|X}(\tilde{x}|x)$ is the conditional probability density of the random variable $\widetilde{X}$ given that $X=x$, and it is given by
\begin{equation}\label{new12eqn3}
    p_{\widetilde{X}|X}(\tilde{x}|x) \d \tilde{x} = \Tr[E_{\tilde{x}}\rho_x] \d \tilde{x}.
\end{equation}
By Theorem~\ref{thm1}, we have
\begin{align}\label{new10eqn2}
 E_{\tilde{x}} \d \tilde{x} &= (2\pi)^{-n} \hat{D}(-\tilde{y}) \sigma \hat{D}(\tilde{y}) \d \tilde{y}
\end{align}
where $\tilde{y}= r_\rho+ JL (\tilde{x}-\mu)$ and $J$ is given by \eqref{new6eqn1}.
This also gives us the relation $\d \tilde{y}=\left|\det [JL]\right| \d \tilde{x}$.
% The relation \eqref{new10eqn4} implies
% \begin{align}
%     \d \tilde{y} &= \left|\det \left[\sqrt{I+ \left(V_\rho \Omega\right)^{-2}} V_\rho \left(V_\rho - V_{\rho_0}\right)^{-1} L \right]\right| \d \tilde{x}\\
%     &= \left|\det \sqrt{I+ \left(V_\rho \Omega\right)^{-2}} \det V_\rho \det [V_\rho - V_{\rho_0}]^{-1} \det L \right| \d \tilde{x} \\
%     &= \left|\sqrt{\det[I-(V_\rho i\Omega)^{-1}] \det[I+(V_\rho i\Omega)^{-1}]} \det V_\rho \det [2L\Sigma L^T]^{-1} \det L \right| \d \tilde{x} \\
%     &= \sqrt{\det[V_\rho]^{-2}\det[(V_\rho- i\Omega)/2] \det[(V_\rho+ i\Omega)/2]} \det V_\rho \det \Sigma^{-1} \d \tilde{x} \\
%     &= \sqrt{\det[(V_\rho - i\Omega)/2] \det[(V_\rho + i\Omega)/2]} \det \Sigma^{-1} \d \tilde{x} \\
%     &= \det[(V_\rho - i\Omega)/2] \det \Sigma^{-1} \d \tilde{x} \label{new10eqn1}.
% \end{align}
By substituting \eqref{new10eqn2} into \eqref{new12eqn3} we thus get
\begin{equation}\label{new13eqn1}
    p_{\widetilde{X}|X}(\tilde{x}|x) \d \tilde{x} = \dfrac{\left|\det [JL]\right|}{(2\pi)^{n}}   \Tr[ \hat{D}(-\tilde{y}) \sigma \hat{D}(\tilde{y})\rho_x] \d \tilde{x}.
\end{equation}
By the overlap formula for Gaussian states \cite[Eq.~(4.51)]{S17} and some simplifying, we  have
\begin{align}\label{new13eqn2}
    \Tr[\hat{D}(-\tilde{y}) \sigma \hat{D}(\tilde{y}) \rho_x] &= \dfrac{1}{\sqrt{\det\left[ (V_{\sigma}+V_{\rho_0})/2  \right]}} \exp\! \left[-\dfrac{1}{2}(\tilde{x}-\mu_x)^T \widetilde{\Sigma}^{-1} (\tilde{x}-\mu_x) \right],
\end{align}
where
% \begin{align}
%     \mu_x & = \mu + L^{-1} \left(V_\rho - V_{\rho_0}\right) V_\rho^{-1} \left(\sqrt{I+\left(V_\rho \Omega\right)^{-2}} \right)^{-1}L (x-\mu), \\
%     \widetilde{\Sigma} &= \dfrac{1}{2} L^{-1} \left(V_\rho - V_{\rho_0}\right) V_\rho^{-1} \left(\sqrt{I-(V_\varrho \Omega)^{-2}} \right)^{-1} (V_{\sigma}+V_{\rho_0}) \left(\sqrt{I-(\Omega V_\varrho)^{-2}} \right)^{-1} \left(V_\rho - V_{\rho_0}\right) V_\rho^{-1} L^T \\
%     &= \Sigma - 4 \Sigma L^T V_\rho^{-1} \left(\sqrt{I-(V_\varrho \Omega)^{-2}} \right)^{-1} L \Sigma L^T \left(\sqrt{I-(\Omega V_\varrho)^{-2}} \right)^{-1}V_\rho^{-1} L\Sigma \label{new12eqn1}. 
% \end{align}
\begin{align}
    \mu_x & =\mu+L^{-1}J^{-1}L(x-\mu), \\
    \widetilde{\Sigma} &= \dfrac{L^{-1}J^{-1}[V_{\sigma}+V_{\rho_0}]J^{-T}L^{-T}}{2}. \label{new12eqn5}
\end{align}
% We used the relations \eqref{new14eqn1} and \eqref{new10eqn3} in \eqref{new12eqn1}.
Substituting \eqref{new13eqn2} into \eqref{new13eqn1} gives 
\begin{align}
    p_{\widetilde{X}|X}(\tilde{x}|x) \d \tilde{x} &=\dfrac{1}{(2\pi)^{n}  \sqrt{\det \widetilde{\Sigma}}} \exp\! \left[-\frac{1}{2}(\tilde{x}-\mu_x)^T \widetilde{\Sigma}^{-1} (\tilde{x}-\mu_x) \right] \d \tilde{x}.
\end{align}
% \begin{align}
%     p(\tilde{x}|x) \d \tilde{x}    &=\dfrac{\det[V_\rho - i\Omega]}{(2\pi)^{n} \det \Sigma \sqrt{\det\left[V_{\rho_0} + V_{\sigma} \right]}}  \exp\! \left[-\frac{1}{2}(\tilde{x}-\mu_x)^T \widetilde{\Sigma}^{-1} (\tilde{x}-\mu_x) \right] \d \tilde{x}.
% \end{align}
Thus, $p_{\widetilde{X}|X}(\tilde{x}|x)$ is a Gaussian probability density function with mean vector $\mu_x$ and covariance matrix~$\widetilde{\Sigma}$. 
Using the formula discussed in Appendix~\ref{gaussianformula}, we get
\begin{align}
    \int_{\mathbb{R}^{2n}}\!\! \d \tilde{x}\, \|x-\tilde{x}\|^2 p_{\widetilde{X}|X}(\tilde{x}|x)  &= \|\mu_x-x\|^2+ \Tr\widetilde{\Sigma} =\left\|\left(L^{-1}J^{-1}L-I \right)(x-\mu)\right\|^2+ \Tr\widetilde{\Sigma}.
\end{align}
Substituting the above value into \eqref{new4eqn6} gives
\begin{align}
    \mathbb{E}[\|X-\widetilde{X}\|^2] &= \int_{\mathbb{R}^{2n}}\!\! \d x \,  \left(\left\|\left(L^{-1}J^{-1}L-I \right)(x-\mu) \right\|^2+ \Tr\widetilde{\Sigma} \right) p(x)  \\
    &=  \int_{\mathbb{R}^{2n}} \!\! \d x\,    \left\|\left(L^{-1}J^{-1}L-I \right)(x-\mu) \right\|^2 p(x)+\Tr\widetilde{\Sigma}.
\end{align}
We note that the matrix $\left(L^{-1}J^{-1}L-I \right)$ is invertible, the proof of which is given in Appendix~\ref{invert1}.
By a change of variable (i.e., $z = \left(L^{-1}J^{-1}L-I \right)(x-\mu)$), we get
\begin{align}
    \mathbb{E}[\|X-\widetilde{X}\|^2] 
    &= \left|\det[L^{-1}J^{-1}L-I]\right|^{-1} \int_{\mathbb{R}^{2n}}\!\! \d z\,    \|z\|^2 p(\mu+[L^{-1}J^{-1}L-I]^{-1}z)+\Tr\widetilde{\Sigma} \\
    &=  \int_{\mathbb{R}^{2n}}\!\! \d z\,    \|z\|^2 q(z)+\Tr\widetilde{\Sigma}, \label{new12eqn4}
\end{align}
where $q(z)$ is the Gaussian probability density function on $\mathbb{R}^{2n}$ with zero mean vector and covariance matrix $\left(L^{-1}J^{-1}L-I \right)\Sigma \left(L^{-1}J^{-1}L-I \right)^T$.
Again, by applying the formula from Appendix~\ref{gaussianformula} to \eqref{new12eqn4}, we get
\begin{align}
    \mathbb{E}[\|X-\widetilde{X}\|^2]     
        &=  \Tr\!\left[\left(L^{-1}J^{-1}L-I \right)\Sigma\left(L^{-1}J^{-1}L-I \right)^T\right]+\Tr\widetilde{\Sigma}. \label{new12eqn7}
\end{align}
We simplify $\widetilde{\Sigma}$ given by \eqref{new12eqn5} using the relations \eqref{new14eqn1} and  \eqref{new10eqn3}  as follows
\begin{align}
    \widetilde{\Sigma} &= L^{-1}J^{-1}\left[-(V_\rho-V_{\rho_0}) +\sqrt{I+\left(V_\rho \Omega \right)^{-2}} V_\rho \left(V_\rho-V_{\rho_0} \right)^{-1} V_\rho \sqrt{I+\left(\Omega V_\rho \right)^{-2}} \right]J^{-T}L^{-T}/2 \\
    &= -L^{-1}J^{-1}L\Sigma L^T J^{-T}L^{-T} + \dfrac{1}{2} L^{-1}J^{-1}\sqrt{I+\left(V_\rho \Omega \right)^{-2}} V_\rho \left(V_\rho-V_{\rho_0} \right)^{-1} V_\rho \sqrt{I+\left(\Omega V_\rho \right)^{-2}} J^{-T}L^{-T} \\
    &= -L^{-1}J^{-1}L\Sigma L^T J^{-T}L^{-T} + \dfrac{1}{2} L^{-1}\left(V_\rho - V_{\rho_0}\right)L^{-T} \\
    &= -L^{-1}J^{-1}L\Sigma J^{-T}L^{-T} + \Sigma. \label{new12eqn6}
\end{align}
By substituting \eqref{new12eqn6} into \eqref{new12eqn7}, we get
\begin{align}
   \mathbb{E}[\|X-\widetilde{X}\|^2]  &=   \Tr\!\left[\left(L^{-1}J^{-1}L-I \right)\Sigma\left(L^{-1}J^{-1}L-I \right)^T\right]+\Tr\!\left[-L^{-1}J^{-1}L\Sigma J^{-T}L^{-T} + \Sigma\right]  \\
    &= 2\Tr\Sigma-2\Tr\!\left[L^{-1}J^{-1}L \Sigma\right] \\
    &= 2 \Tr\!\left[(I-L^{-1}J^{-1}L)\Sigma\right]. \label{new12eqn8}
\end{align}
We obtain the desired expression \eqref{new10eqn6} by resubstituting the value of $J$ in \eqref{new12eqn8}.

%%%%%%%%%%%%%%%%%%%%%%%%%%%%%%%%%%%%%%%%%%%%%%%%%%%%%%%%%%%%%%%%%%
%\section{Simplification of $J(V_\rho+V_4)J^T$}\label{app_b}
% From \eqref{new4eqn1}, we have
% \begin{align}\label{new4eqn12}
%     W_{\sigma}+W_\rho &= \sqrt{I-W_\rho^{-2}} W_\rho (W_\rho-W_{\rho_0})^{-1}W_\rho \sqrt{I-W_\rho^{-2}}.
% \end{align}

\section{Proof of Theorem~\ref{thm3}}\label{proof_pgi}
Let $\tau$ be a generic faithful Gaussian state. 
We have
\begin{align}\label{new20eqn2}
    \rho_x^{1/2} \rho^{-1/2} \tau \rho^{-1/2} \rho_x^{1/2} &= \dfrac{Z_\rho}{Z_{\rho_0} Z_\tau} \exp[-\hat{H}_{\rho_x}/2] \exp[\hat{H}_{\rho}/2]\exp[-\hat{H}_{\tau}] \exp[\hat{H}_{\rho}/2] \exp[-\hat{H}_{\rho_x}/2].
\end{align}
We first combine the product of exponential operators $\exp[\hat{H}_{\rho}/2]\exp[-\hat{H}_{\tau}] \exp[\hat{H}_{\rho}/2]$.
The golden rule implies that there exists an operator $\hat{H}_5$ of the form
\begin{align}\label{new2eqn1}
    \hat{H}_5 = \dfrac{i}{2} \hat{r}^T \Omega X_5 \hat{r}+i s_5^T \Omega \hat{r}+\dfrac{i}{2} a_5,
\end{align}
that satisfies
\begin{align}
    \exp[\hat{H}_{\rho}/2]\exp[-\hat{H}_{\tau}] \exp[\hat{H}_{\rho}/2]=\exp[\hat{H}_5].\label{new22eqn3}
\end{align}
% Let $M_5$ be the matrix corresponding to $\hat{H}_5$, given by
% \begin{align}
%     M_5=\begin{bmatrix}0 & z_5^T \Omega^T & b_5 \\ 0 & Y_5 & z_5 \\ 0 & 0 & 0 \end{bmatrix}.
% \end{align}
By similar arguments as in the proof of Theorem~\ref{thm1}, we get $X_5=-i\Omega H_5$ and $s_5 = X_5 y_5$, where 
\begin{align}
    H_5 &=2i\Omega \arcoth(V_5i\Omega),\\
    V_5 &\coloneqq -V_\rho + \sqrt{I+\left(V_\rho \Omega\right)^{-2}} V_\rho \left(V_\rho-V_\tau\right)^{-1} V_\rho \sqrt{I+\left(\Omega V_\rho\right)^{-2}},\label{new22eqn1}\\
    y_5&=r_{\rho}+J_5(r_\tau-r_\rho),\label{new22eqn7}\\
    J_5 &\coloneqq \sqrt{I+\left(V_\rho \Omega\right)^{-2}}V_\rho\left(V_\rho-V_\tau\right)^{-1}.\label{new22eqn11}
\end{align}
We note that $V_5-i\Omega >0$, whenever $V_\tau < V_\sigma$. Indeed,
by using the relation \eqref{new3eqn10}, we get
\begin{align}
    V_5-i\Omega 
        &= -(V_\rho+i\Omega) + \sqrt{I+\left(V_\rho \Omega\right)^{-2}} V_\rho \left(V_\rho-V_\tau\right)^{-1} V_\rho \sqrt{I+\left(\Omega V_\rho\right)^{-2}}  \\
        &= \sqrt{I+\left(V_\rho \Omega\right)^{-2}} V_\rho \left[\left(V_\rho-V_\tau\right)^{-1}- \left(V_\rho-i\Omega\right)^{-1} \right] V_\rho \sqrt{I+\left(\Omega V_\rho\right)^{-2}}.
\end{align}
Since $V_\tau < V_\sigma$ implies $\left(V_\rho-V_\tau\right)^{-1}- \left(V_\rho-i\Omega\right)^{-1} > 0$, we have $V_5-i\Omega >0$. The fact that $V_5$ is a real matrix such that $V_5-i\Omega >0$ implies $V_5 >0$, a proof of which can be found on Page~58 of \cite{S17}. Thus, $V_5$ is a legitimate covariance matrix of a faithful Gaussian state.  
 It is also easy to see that $y_5$ is a real vector.
% Also, by following similar arguments given in the proof of Proposition~12 of \cite{seshadreesan2018renyi}, we get
% \begin{align}\label{new20eqn13}
%     \det(\left[V_5+i\Omega\right]/2) 
%         &= \dfrac{Z_\rho^2 Z_\tau^2}{\det([V_\rho-V_\tau]/2)}.
% \end{align}
Now, substituting the values of $X_5$ and $s_5$ in \eqref{new2eqn1} gives
\begin{align}
    \hat{H}_5 &= -\frac{1}{2} \left(\hat{r}-y_5 \right)^T H_5 \left(\hat{r}-y_5\right) + \dfrac{1}{2} \left( ia_5+y_5^T H_5y_5 \right).\label{new22eqn2}
\end{align}
From \eqref{new20eqn2}, using the relations \eqref{new22eqn3} and \eqref{new22eqn2}, we have
\begin{multline}\label{new2eqn3}
     \rho_x^{1/2} \rho^{-1/2} \tau \rho^{-1/2} \rho_x^{1/2}= \dfrac{Z_\rho }{Z_{\rho_0} Z_\tau}  \exp\!\left[\frac{1}{2}\left( ia_5+y_5^T H_5y_5 \right)\right] \\ \exp[-\hat{H}_{\rho_x}/2]\exp\!\left[-\frac{1}{2} \left(\hat{r}-y_5 \right)^T H_5 \left(\hat{r}-y_5\right)\right]\exp[-\hat{H}_{\rho_x}/2].
\end{multline}
Again, by applying the golden rule and arguments similar to those given in the proof of Theorem~\ref{thm1}, we get an operator $\hat{H}_6$ of the form
\begin{align}\label{new2eqn2}
    \hat{H}_6 = \dfrac{i}{2} \hat{r}^T \Omega X_6 \hat{r}+i s_6^T \Omega \hat{r}+\dfrac{i}{2} a_6,
\end{align}
 satisfying
\begin{align}\label{new2eqn5}
    \exp[-\hat{H}_{\rho_x}/2]\exp\!\left[-\frac{1}{2} \left(\hat{r}-y_5 \right)^T H_5 \left(\hat{r}-y_5\right)\right]\exp[-\hat{H}_{\rho_x}/2]=\exp[\hat{H}_6].
\end{align}
Moreover, we have $X_6=-i\Omega H_6$ and $s_6 = X_6 y_6$, where 
\begin{align}
    H_6 &=2i\Omega \arcoth(V_6i\Omega),\\
    V_6 &\coloneqq V_{\rho_0} - \sqrt{I+\left(V_{\rho_0} \Omega\right)^{-2}} V_{\rho_0} \left(V_5+V_{\rho_0}\right)^{-1} V_{\rho_0} \sqrt{I+\left(\Omega V_{\rho_0}\right)^{-2}},\label{new25eqn7}\\
    y_6&=r_{\rho_x}+J_6(y_5-r_{\rho_x}),\label{new21eqn1}\\
    J_6 &\coloneqq \sqrt{I+\left(V_{\rho_0}\Omega\right)^{-2}}V_{\rho_0} \left(V_5+V_{\rho_0}\right)^{-1}.\label{new22eqn6}
\end{align}

The following argument shows that $V_6$ is a legitimate covariance matrix for a faithful Gaussian state if $V_\tau < V_\rho$. 
By \eqref{new3eqn10}, we have
\begin{align}
    V_6-i\Omega &=  \sqrt{I+\left(V_{\rho_0} \Omega\right)^{-2}} V_{\rho_0} \left[\left(V_{\rho_0}+i\Omega \right)^{-1}-\left(V_5+V_{\rho_0}\right)^{-1}\right] V_{\rho_0} \sqrt{I+\left(\Omega V_{\rho_0}\right)^{-2}}.
\end{align}
If $V_\tau < V_\rho$, we know that $V_5+i\Omega > 0$.  This implies $\left(V_{\rho_0}+i\Omega \right)^{-1}-\left(V_5+V_{\rho_0}\right)^{-1}>0$. We thus get $V_6-i\Omega >0$.
We also note that $y_6$ is a real vector since $y_5$ is a real vector and $J_6$ is a real matrix. 
% Again, by following similar arguments given in the proof of Proposition~12 of \cite{seshadreesan2018renyi}, and then using \eqref{new20eqn13}, we get
% \begin{align}
%     \det([V_6+i\Omega]/2) 
%         &= \dfrac{Z_{\rho_0}^2 Z_\rho^2 Z_\tau^2}{\det([V_5+V_{\rho_0}]/2)\det([V_\rho-V_\tau]/2)}. \label{new20eqn14}
% \end{align}
Substituting the values of $X_6$ and $s_6$ into \eqref{new2eqn2} gives
\begin{align}\label{new2eqn7}
    \hat{H}_6 &= -\frac{1}{2} \left(\hat{r}-y_6 \right)^T H_6 \left(\hat{r}-y_6\right) + \dfrac{1}{2} \left( ia_6+y_6^T H_6y_6 \right).
\end{align}
From \eqref{new2eqn3}, using the relations \eqref{new2eqn5} and \eqref{new2eqn7}, we get
\begin{align}\label{new20eqn10}
    \rho_x^{1/2} \rho^{-1/2} \tau \rho^{-1/2} \rho_x^{1/2} &= \dfrac{\tilde{\xi} Z_\rho }{Z_{\rho_0} Z_\tau}   \exp\!\left[-\frac{1}{2} \left(\hat{r}-y_6 \right)^T H_6 \left(\hat{r}-y_6\right)\right],
\end{align}
where $\tilde{\xi}$ is a scalar given by
\begin{align}
    \tilde{\xi} \coloneqq \exp\!\left[\frac{1}{2}\left( i(a_5+a_6)+y_5^T H_5y_5 +y_6^T H_6y_6 \right)\right].
\end{align}

In what follows, we will simplify \eqref{new20eqn10} to get the desired structure of the instrument. 
By \eqref{new20eqn10}, we get
\begin{align}
    \Tr\!\left[p(x)\rho_x^{1/2} \rho^{-1/2} \tau \rho^{-1/2} \rho_x^{1/2} \right] \d x
        &= \dfrac{\tilde{\xi} p(x) Z_\rho }{Z_{\rho_0} Z_\tau}   \Tr\exp\!\left[-\frac{1}{2} \left(\hat{r}-y_6 \right)^T H_6 \left(\hat{r}-y_6\right)\right] \d x \\
        &= \dfrac{\tilde{\xi} p(x) Z_\rho }{Z_{\rho_0} Z_\tau} \sqrt{\det([V_6+i\Omega]/2)} \d x.\label{new20eqn12}
\end{align}
Also, we have
\begin{align}
    & \Tr\!\left[p(x)\rho_x^{1/2} \rho^{-1/2}  \tau \rho^{-1/2} \rho_x^{1/2} \right] \d x \nonumber \\
        &\hspace{1cm}= \Tr\!\left[ \tau  \rho^{-1/2} p(x) \rho_x \rho^{-1/2} \right] \d x \\
        &\hspace{1cm}= \dfrac{1}{(2\pi)^n} \Tr\!\left[ \tau   \hat{D}(-y) \sigma \hat{D}(y) \right] \d y \\
        &\hspace{1cm}= \dfrac{(2\pi)^{-n}}{ \sqrt{\det([V_\tau + V_{\sigma}]/2)}} \exp\!\left[-\frac{1}{2}(y-r_\tau)^T ([V_\tau + V_{\sigma}]/2)^{-1} (y-r_\tau)\right] \d y,\label{new20eqn11}
\end{align}
where $y$ and  $\sigma$ are given as in Theorem~\ref{thm1}.
The first equality follows from the cyclic property of trace, the second equality follows from \eqref{new3eqn15}, and the third equality follows from the overlap formula for Gaussian states \cite[Eq.~(4.51)]{S17}.
By comparing the two expressions \eqref{new20eqn11} and \eqref{new20eqn12}, we get
\begin{align}
    \dfrac{\tilde{\xi} p(x) Z_\rho }{Z_{\rho_0} Z_\tau}  \d x 
        = \dfrac{(2\pi)^{-n}}{\sqrt{\det([V_6+i\Omega]/2)} \sqrt{\det([V_\tau + V_{\sigma}]/2)}} \exp\!\left[-\frac{1}{2}(y-r_\tau)^T ([V_\tau + V_{\sigma}]/2)^{-1} (y-r_\tau)\right] \d y.
\end{align}
Substituting the above value into \eqref{new20eqn10} gives
\begin{align}\label{new21eqn2}
    p(x)\rho_x^{1/2} \rho^{-1/2} \tau \rho^{-1/2} \rho_x^{1/2} \d x
        &= \dfrac{\tilde{p}(y)}{\sqrt{\det([V_6+i\Omega]/2)}}\exp\!\left[-\frac{1}{2} \left(\hat{r}-y_6 \right)^T H_6 \left(\hat{r}-y_6\right)\right] \d y.
\end{align}
% Let $\rho_6(y_6, V_6)$ be the Gaussian state with mean vector $y_6$ and covariance matrix $V_6$. We thus have
% \begin{align}
%      p(x)\rho_x^{1/2} \rho^{-1/2} \tau \rho^{-1/2} \rho_x^{1/2} 
%         &= \dfrac{(2\pi)^{-n} \det J^{-1} }{ \sqrt{\det([V_\tau + V_{\sigma}]/2)}} \nonumber\\
%         &\hspace{0.5cm} \exp\!\left[-(r_\tau-y)^T (V_\tau+V_{\sigma})^{-1} (r_\tau-y)\right]\rho_6(y_6, V_6).
% \end{align}
% Substitute the value of $\det([V_6+i\Omega]/2)$ from \eqref{new20eqn14} to get
% \begin{align}
%     p(x)\rho_x^{1/2} \rho^{-1/2} \tau \rho^{-1/2} \rho_x^{1/2} 
%         &= \dfrac{(2\pi)^{-n} \det J^{-1} \sqrt{\det([V_5+V_{\rho_0}]/2)\det([V_\rho-V_\tau]/2)} }{Z_{\rho_0} Z_\rho Z_\tau \sqrt{\det([V_\tau + V_{\sigma}]/2)}} \nonumber\\
%         &\hspace{0.5cm} \exp\!\left[-(r_\tau-y)^T (V_\tau+V_{\sigma})^{-1} (r_\tau-y)\right] \exp\!\left[-\frac{1}{2} \left(\hat{r}-y_6 \right)^T H_6 \left(\hat{r}-y_6\right)\right] \\
%         &= \dfrac{\tilde{p}(y) \sqrt{\det([V_5+V_{\rho_0}]/2)\det([V_\rho-V_\tau]/2)} }{Z_{\rho_0} Z_\rho Z_\tau \det J}
%         \exp\!\left[-\frac{1}{2} \left(\hat{r}-y_6 \right)^T H_6 \left(\hat{r}-y_6\right)\right],
% \end{align}
where $\tilde{p}(y)$ is the Gaussian probability density function on $\mathbb{R}^{2n}$ with the mean vector $r_\tau$ and the covariance matrix $(V_\tau+V_{\sigma})/2$.
% \begin{align}
%     \tilde{p}(y) &= \dfrac{1}{(2\pi)^n \sqrt{\det([V_\tau + V_{\sigma}]/2)}} \exp\!\left[-\frac{1}{2}(r_\tau-y)^T ([V_\tau + V_{\sigma}]/2)^{-1} (r_\tau-y)\right].
% \end{align}
In order to express the exponential operator in the right-hand side of \eqref{new21eqn2} as a function of $y$, we express $y_6$ in terms of $y$. Recall from \eqref{new20eqn9} that $x=L^{-1}J^{-1}(y-r_\rho)+\mu$, where $J$ is given by \eqref{new6eqn1}.
So, using the relations $r_{\rho_x}=r_{\rho_0}+Lx$, $r_{\rho}=r_{\rho_0}+L\mu$, and \eqref{new21eqn1}, we get
\begin{align}
    y_6 &= (I-J_6) r_{\rho_x} + J_6y_5 \\
        &= (I-J_6) (Lx+r_{\rho_0}) + J_6y_5 \\
        &= (I-J_6) \left[J^{-1}(y-r_{\rho})+L\mu+r_{\rho_0}\right] + J_6y_5 \\
        &= (I-J_6) J^{-1}y -(I-J_6)J^{-1}r_\rho + (I-J_6)r_{\rho}  + J_6y_5 \\
        &=  (I-J_6) J^{-1}y -(I-J_6)J^{-1}r_\rho + r_{\rho}  + J_6(y_5-y_\rho).
\end{align}
By substituting the value of $y_5-r_\rho$ from \eqref{new22eqn7} in the above relation, we get
\begin{align}
     y_6 
        &= (I-J_6) J^{-1}y +\left[I -(I-J_6)J^{-1}\right]r_\rho + J_6J_5 (r_\tau-r_\rho)
        \coloneqq J_7(y- r_\rho)+y_7,\label{new22eqn8}
\end{align}
% \begin{align}
%     y_6 &= (I-J_6^{-1}) r_{\rho_x} + J_6^{-1}y_5 \\
%         &= (I-J_6^{-1}) (Lx+r_{\rho_0}) + J_6^{-1}y_5 \\
%         &= (I-J_6^{-1}) \left(2L\Sigma L^T V_\rho^{-1}\left(\sqrt{I+\left(V_\rho \Omega\right)^{-2}}\right)^{-1}(y-r_{\rho})+r_{\rho}\right) + J_6^{-1}y_5 \\
%         &= 2(I-J_6^{-1}) L\Sigma L^T V_\rho^{-1}\left(\sqrt{I+\left(V_\rho \Omega\right)^{-2}}\right)^{-1} y  + (I-J_6^{-1})\left(I-2L\Sigma L^T V_\rho^{-1}\left(\sqrt{I+\left(V_\rho \Omega\right)^{-2}}\right)^{-1} \right)r_\rho + J_6^{-1}y_5.
% \end{align}
% The first equality follows from \eqref{new21eqn1}, the second equality follows from the relation $r_{\rho_x}=Lx+r_{\rho_0}$, the third equality follows from \eqref{new20eqn9}, and the fourth equality uses the relation \eqref{new14eqn1}.
where $J_7$ is a real matrix and $y_7$ is a real vector, given by
\begin{align}
    J_7 &= (I-J_6) J^{-1} =2\left(I-J_6\right) L\Sigma L^T V_\rho^{-1}\left(\sqrt{I+\left(V_\rho \Omega\right)^{-2}}\right)^{-1}, \label{new22eqn5} \\
    y_7 &= r_\rho + J_6J_5 (r_\tau-r_\rho).\label{new22eqn10}
\end{align}
From \eqref{new21eqn2} and \eqref{new22eqn8}, we thus get
\begin{align}\label{new21eqn3}
    p(x)\rho_x^{1/2} \rho^{-1/2} \tau \rho^{-1/2} \rho_x^{1/2} \d x
        &= \tilde{p}(y)\hat{D}(-J_7(y-r_\rho)) \rho_7 \hat{D}(J_7(y-r_\rho)) \d y,
\end{align}
where $\rho_7$ is a faithful Gaussian state with mean $r_{\rho_7}=y_7$ given by \eqref{new22eqn10} and covariance matrix $V_{\rho_7}=V_6$ given by \eqref{new25eqn7}. 
We note that $J_7$ is invertible, which is shown in Appendix~\ref{invert2}.
Therefore, a change of variable on the right-hand side of \eqref{new21eqn3} (i.e., $z=J_7(y-r_\rho)$) gives
\begin{align}\label{new23eqn1}
    p(x)\rho_x^{1/2} \rho^{-1/2} \tau \rho^{-1/2} \rho_x^{1/2} \d x
        &= \tilde{q}(z)\hat{D}(-z) \rho_7 \hat{D}(z) \d z,
\end{align}
where $\tilde{q}(z)$ is a Gaussian density function with mean vector $J_7(r_\tau-r_\rho)$ and covariance matrix $J_7(V_\tau+V_{\sigma})J_7^T/2$. The variable $z$ is related to $x$ by
\begin{align}
    z=J_7(y-r_\rho)
    &=J_7\sqrt{I+\left(V_\rho \Omega\right)^{-2}} V_\rho(V_\rho-V_{\rho_0})^{-1}L(x-\mu) \\
    &=J_7\sqrt{I+\left(V_\rho \Omega\right)^{-2}} V_\rho L^{-T}\Sigma^{-1}(x-\mu)/2.
\end{align}
The post-measurement state of the instrument is then obtained from \eqref{new23eqn1}:  
\begin{align}
    \dfrac{p(x)\rho_x^{1/2} \rho^{-1/2} \tau \rho^{-1/2} \rho_x^{1/2}}{\Tr\!\left[p(x)\rho_x^{1/2} \rho^{-1/2} \tau \rho^{-1/2} \rho_x^{1/2}\right]}&= \hat{D}(-z) \rho_7 \hat{D}(z).
\end{align}

To obtain the expected output state $\int_{\mathbb{R}^{2n}}\!\! \d x \ p(x)\rho_x^{1/2} \rho^{-1/2} \tau \rho^{-1/2} \rho_x^{1/2}$ that results from discarding the measurement outcome, consider the following argument. By integrating on both sides of \eqref{new23eqn1}, and using the classical mixing formula given in \cite[Eq.~(5.76)]{S17}, we get
\begin{align}
    \int_{\mathbb{R}^{2n}}\!\! \d x \ p(x)\rho_x^{1/2} \rho^{-1/2} \tau \rho^{-1/2} \rho_x^{1/2}
        &=\int_{\mathbb{R}^{2n}}\!\! \d z \ \tilde{q}(z)\hat{D}(-z) \rho_7 \hat{D}(z) = \tilde{\tau},
\end{align}
where $\tilde{\tau}$ is a Gaussian state with mean vector and covariance matrix given by
\begin{align}
    r_{\tilde{\tau}}&= r_{\rho_7}+ J_7(r_\tau-r_\rho), \\
    V_{\tilde{\tau}}&= V_{\rho_7}+J_7 \left[  V_\tau + V_{\sigma}\right]J_7^T.
\end{align}

%%%%%%%%%%%%%%%%%%%%%%%%%%%%%%%%%%%%%%%%%%%%%%%%%%%%%%%%%%%%%%%%%%
\section{Invertibility of the matrix $L^{-1}J^{-1}L-I$}

\label{invert1}

We show that the matrix $L^{-1}J^{-1}L-I$ is invertible, where $L$ is defined in \eqref{neweqn4} and $J$ in \eqref{new6eqn1}. Recall that it is assumed that $L$ is invertible and $J$ is invertible because it is a product of invertible matrices. Since
$L^{-1}J^{-1}L-I=L^{-1}J^{-1}(I-J)L,$
it suffices to show that $I-J$ is invertible. We repeatedly make use of the functional calculus of matrices: if $A$ is a diagonalizable matrix and $X$ is an invertible matrix, then $f(X^{-1}AX)=X^{-1}f(A)X$ holds for any continuous function $f$ defined on the spectrum of $A$.

From \eqref{new6eqn1}, we get
\begin{align}
    I-J 
        &= \left[V_\rho-V_{\rho_0}-\sqrt{I+\left(V_\rho \Omega\right)^{-2}} V_\rho\right] (V_\rho-V_{\rho_0})^{-1} \\
        &=  \left[\left(I-\sqrt{I+\left(V_\rho \Omega\right)^{-2}} \right)V_\rho -V_{\rho_0}\right] (V_\rho-V_{\rho_0})^{-1},
\end{align}
which gives
\begin{align}\label{iminusj_invert}
    (I-J) (V_\rho-V_{\rho_0})
        &=  \left(I-\sqrt{I+\left(V_\rho \Omega\right)^{-2}} \right)V_\rho -V_{\rho_0}.
\end{align}
By Williamson's theorem \cite{williamson1936algebraic}, there exists a symplectic matrix $S$, which by definition satisfies $S^T \Omega S=\Omega$, such that $V_\rho  = S^T DS$, where $D=\Lambda \otimes I_2$, and $\Lambda$ is a diagonal matrix with positive diagonal elements. We refer to Refs.~\cite{folland1989harmonic, simon1999congruences, parthasarathy2012symmetry} for more details on Williamson's theorem.
From \eqref{iminusj_invert}, and using the fact $S \Omega = \Omega S^{-T}$, we thus get
\begin{align}
    (I-J) (V_\rho-V_{\rho_0})
        &=  \left(I-\sqrt{I+(S^T D S \Omega)^{-2}} \right)S^T D S -V_{\rho_0} \label{iminusj_invert_1} \\
        &=\left(I-\sqrt{I+(S^T D\Omega S^{-T})^{-2}} \right)S^T D S -V_{\rho_0} \\
        &=S^T \left(I-\sqrt{I+(D\Omega)^{-2}} \right)S^{-T} S^T D S -V_{\rho_0} \\
        &=S^T \left(I-\sqrt{I-D^{-2}} \right) D S -V_{\rho_0}. \label{iminusj_invert_2}
        \end{align}
We note here that $D> I_n$. Indeed, by applying Williamson's theorem in condition \eqref{faithful_condition}, we get $D+i\Omega >0$, which is equivalent to $D > I_n$. We also have the function relation, for $x > 1$,
\begin{align}
     \left(1-\sqrt{1-x^{-2}} \right) x &= \dfrac{x^{-1}}{1+\sqrt{1-x^{-2}}} < x^{-1}.
\end{align}
Using the above function relation in \eqref{iminusj_invert_2}, and then simplifying further, gives
\begin{align}
    (I-J) (V_\rho-V_{\rho_0})
        &< S^T D^{-1} S -V_{\rho_0} \\
        &= \left(S^{-1} D S^{-T}\right)^{-1} -V_{\rho_0} \\
        &= \left( S^{-1} \Omega D \Omega^T S^{-T} \right)^{-1} -V_{\rho_0} \\
        &= \left( \Omega S^{T}  D S \Omega^T \right)^{-1} -V_{\rho_0} \\
        &= \left( \Omega V_\rho \Omega^T \right)^{-1} -V_{\rho_0} \\
        &= \Omega V_\rho^{-1} \Omega^T -V_{\rho_0}. \label{iminusj_invert_3}
\end{align}
Recall that $V_\rho > V_{\rho_0}$, which implies $V_\rho^{-1} < V_{\rho_0}^{-1}$. From \eqref{iminusj_invert_3} we thus get
\begin{align}
    (I-J) (V_\rho-V_{\rho_0})
        &<  \Omega V_{\rho_0}^{-1} \Omega^T -V_{\rho_0}.
\end{align}
The matrix $\Omega V_{\rho_0}^{-1} \Omega^T -V_{\rho_0}$ is negative definite, which follows from Lemma~11 of \cite{lami2017log}. We have thus proved that $(I-J) (V_\rho-V_{\rho_0}) < 0$. In particular, $I-J$ is an invertible matrix.

%%%%%%%%%%%%%%%%%%%%%%%%%%%%%%%%%%%%%%%%%%%%%%%%%%%%%%%%%%%%%%%%%%
\section{Invertibility of the matrix $J_7$ from Eq.~(\ref{new25eqn5})}

\label{invert2}
To prove that $J_7=2(I-J_6)L\Sigma L^T V_\rho^{-1}\left(\sqrt{I+\left(V_\rho \Omega\right)^{-2}}\right)^{-1}$ from \eqref{new25eqn5} is invertible, it suffices to show that $I-J_6$ is invertible, where $J_6$ is defined in \eqref{new25eqn4}. 
% The proof below is similar to the proof presented in Appendix~\ref{invert1}.
We have that
\begin{align}
    (I-J_6)\left(V_5+V_{\rho_0}\right)  
        =  V_5+V_{\rho_0}-\sqrt{I+\left(V_{\rho_0}\Omega\right)^{-2}}V_{\rho_0}
        >V_{\rho_0}-\sqrt{I+\left(V_{\rho_0}\Omega\right)^{-2}}V_{\rho_0}.\label{iminusj6_invert_1}
\end{align}
The last inequality follows because $V_5 > 0$, under the given condition $V_\rho > V_\tau$. Apply Williamson's theorem again (as in Appendix~\ref{invert1}) to get a symplectic matrix $S_0$ such that $V_{\rho_0}  = S_0^T D_0S_0$, where $D_0=\Lambda_0 \otimes I_2$, and $\Lambda_0$ is a diagonal matrix with positive diagonal elements.
By applying similar arguments to \eqref{iminusj6_invert_1} as developed in \eqref{iminusj_invert_1}--\eqref{iminusj_invert_2}, we get 
\begin{align}
    (I-J_6)\left(V_5+V_{\rho_0}\right)  
        >S_0^T \left(I-\sqrt{I-D_0^{-2}} \right)D_0S_0>0.\label{iminusj6_invert_2}
\end{align}
We have thus proved that $(I-J_6)\left(V_5+V_{\rho_0}\right)$ is a positive definite matrix. In particular, $I-J_6$ is invertible.

\end{appendices}

\end{document}